\documentclass[12pt]{article}
\usepackage{amsfonts,amsthm,amsmath,amssymb,upgreek}
\usepackage{hyperref}
\usepackage[paper=letterpaper,margin=1in]{geometry}
\usepackage{graphicx}
\usepackage{units}
\usepackage{color}

\def\Tr{{\rm Tr }}
\def\hat{\widehat}
\newcommand{\be}{\begin{equation}}
\newcommand{\ee}{\end{equation}}
\def\t{\tilde}

\def\b{\overline}

\def\t{\widetilde{t}}
\def\T{\widetilde{T}}
\def\db{\delta\hspace{-1.3pt}\beta}
\def\dE{\delta\hspace{-1.3pt}E}
\def\b{\text{\boldmath $\beta$}}
\def\DeltaE{\Delta\hspace{-1pt}E}
\def\Deltat{\Delta}

\begin{document}
\thispagestyle{empty}

\vspace*{.5cm}
\begin{center}

{\bf {\LARGE A semiclassical ramp in SYK and in gravity}\\
\vspace{1cm}}

  \begin{center}

 {\bf Phil Saad,$^{a}$ Stephen H. Shenker,$^{a}$ and Douglas Stanford$^b$}\\
  \bigskip \rm
  
\bigskip
 $^a$ Stanford Institute for Theoretical Physics,\\Stanford University, Stanford, CA 94305\\ \vspace{5pt}
   $^b$Institute for Advanced Study, Princeton, NJ 08540

\rm
  \end{center}

\vspace{1.5cm}
{\bf Abstract}
\end{center}
\begin{quotation}
\noindent

In finite entropy systems, real-time partition functions do not decay to zero at late time. Instead, assuming random matrix universality, suitable averages exhibit a growing ``ramp'' and ``plateau'' structure. Deriving this non-decaying behavior in a large $N$ collective field description is a challenge related to one version of the black hole information problem. We describe a candidate semiclassical explanation of the ramp for the SYK model and for black holes. In SYK, this is a two-replica nonperturbative saddle point for the large $N$ collective fields, with zero action and a compact zero mode that leads to a linearly growing ramp. In the black hole context, the solution is a two-sided black hole that is periodically identified under a Killing time translation. We discuss but do not resolve some puzzles that arise.
\end{quotation}

\vspace{1cm}

\begin{center}
{\it Dedicated to the memory of Joe Polchinski}
\end{center}
\setcounter{page}{0}
\setcounter{tocdepth}{2}
\setcounter{footnote}{0}
\newpage

\tableofcontents

\pagebreak

\section{Introduction}

\subsection{Review and context}

The black hole information problem\footnote{For a review as of 2016 see \cite{Polchinski:2016hrw}.}
is rooted in a deep tension between the smoothness of spacetime geometry and the discreteness of quantum Hilbert space.
A simple example of this tension was described by Maldacena \cite{Maldacena:2001kr}. Consider the correlation function of two simple bulk operators separated in time  a finite distance above the black hole horizon. In a gravity analysis, such correlators exponentially decay\footnote{There are usually small power law tails due to conserved quantities if the operators are local in space.} because of the relaxational behavior of the quasinormal modes of a black hole. This conflicts with a basic fact that in a finite quantum system, correlation functions cannot decay to zero at late time.

One can make the discussion sharp using the boundary description (via AdS/CFT) of a large but finite-entropy AdS black hole. There the observable just described becomes a conventional thermal correlator, e.g.
\be\label{thermalcorr}
f_\beta(T) = \frac{1}{Z}\text{Tr}\left(e^{-\frac{\beta}{2}H}O(T)e^{-\frac{\beta}{2}H}O(0)\right)= \frac{1}{Z}\sum_{n,m} e^{-(\frac{\beta}{2}+iT)E_n}e^{-(\frac{\beta}{2}-iT)E_m}  |\langle n | O |m\rangle|^2.
\ee
Here $O$ is a simple Hermitian operator and $Z$ is the partition function $Z(\beta) \equiv \Tr \exp(-\beta H)$, and the double sum is over the discrete spectrum of energy eigenstates. At early times, such correlators typically do decay exponentially, because of chaotic thermalization.  This is the boundary dual of quasinormal mode relaxation \cite{Horowitz:1999jd}. However, discrete sums of oscillating terms like $f(T)$ cannot decay to zero at late time \cite{Maldacena:2001kr,Dyson:2002pf,Goheer:2002vf,Barbon:2003aq}. Instead, they typically become exponentially small (in the entropy) and fluctuate erratically in time. This effect appears to be invisible in perturbative corrections to classical gravity and so explaining it from the bulk point of view could shed some light on the tension described above.  
 
We expect the squared matrix elements in \eqref{thermalcorr} to vary smoothly because of the Eigenstate Thermalization Hypothesis \cite{deutsch1991quantum,srednicki1994chaos}.  The oscillating phases are the key actors.  To isolate their effect it is useful to define a simpler observable \cite{Papadodimas:2015xma}\footnote{Operators depending only on phases are also discussed in \cite{Susskind:2013lpa} .} where we strip off the matrix elements and relabel $\beta \rightarrow 2\beta$:
\be\label{sff}
Z(\beta+iT) Z(\beta-iT) = \sum_{n,m} e^{- ( \beta+iT)E_n}e^{-(\beta -iT)E_m} = \sum_{m,n}e^{-\beta(E_m + E_n) + iT(E_m -E_n)}.
\ee
People working on quantum chaos have long studied the $\beta = 0$ case of (\ref{sff}), calling it the ``spectral form factor," a name we will often use \cite{haake2010quantum}. At short times $|Z(\beta +iT)|^2 = Z(\beta)^2$. At long times, after doing a bit of time averaging to smooth out the erratic fluctuations, all the terms in $|Z(\beta +iT)|^2$ where $E_n \neq  E_m$ average to zero.  Assuming a generic chaotic spectrum this only leaves the diagonal $n = m$ terms, giving  $|Z(\beta +iT)|^2 = Z(2 \beta)$.  We call this late time region the ``plateau." It is helpful to have in mind a rough picture where we ignore the energy term in the free energy, so the late-time value is $Z(2\beta) = e^{S}$. For systems with many degrees of freedom, this is large but exponentially smaller than the early time value $Z(\beta)^2 = e^{2S}$. The  behavior as one evolves from early to late times will be the focus of this paper.  

A direct calculation of the spectral form factor for long times in theories with a well understood gravity dual (like Super Yang-Mills theory) is currently impossible.  We do not have fine enough control over the detailed structure of the high energy spectrum.    But in recent years a model has been introduced that is tractable, chaotic, and has some aspects of a gravity dual.   This is the Sachdev-Ye-Kitaev (SYK) model \cite{Sachdev:1992fk,KitaevTalks,Kitaev:2017awl}. 

The SYK model is a quantum mechanical model of $N$ interacting Majorana fermions $\psi_{a}$, coupled in sets of $q$ by random couplings $J_{a_1\dots a_q}$,
\be
H = i^{\frac{q}{2}}\sum_{a_1<\dots<a_q}J_{a_1\dots a_q}\psi_{a_1}\dots \psi_{a_q}.
\ee
Properties of the model after averaging over the $J$ ensemble are particularly simple to analyze.  They can be expressed exactly in terms of bilocal collective fields called $G(t,t')$ and $\Sigma(t,t')$. The field $\Sigma(t,t')$ is a Lagrange multiplier that enforces the identification  $G(t,t') = \frac{1}{N} \sum_a \psi_a(t) \psi_a(t')$.  At large $N$ the path integral is described by a saddle point configuration of $G, \Sigma$ and fluctuations around it.   At low temperatures the model is almost conformal. The most important fluctuations in the $G,\Sigma$ fields consist of a soft mode that is described by the Schwarzian theory \cite{KitaevTalks,Maldacena:2016hyu,Kitaev:2017awl}.  In turn, this theory is equivalent \cite{Maldacena:2016upp,Jensen:2016pah,Engelsoy:2016xyb} to an action describing two dimensional dilaton gravity, the Jackiw-Teitelboim (JT) model \cite{Jackiw:1984je,Teitelboim:1983ux,Almheiri:2014cka}.  This provides the basis for connections between AdS${}_2$ black hole physics and the SYK model.

\begin{figure}[ht]
  \centering
  \includegraphics[width=0.6\textwidth]{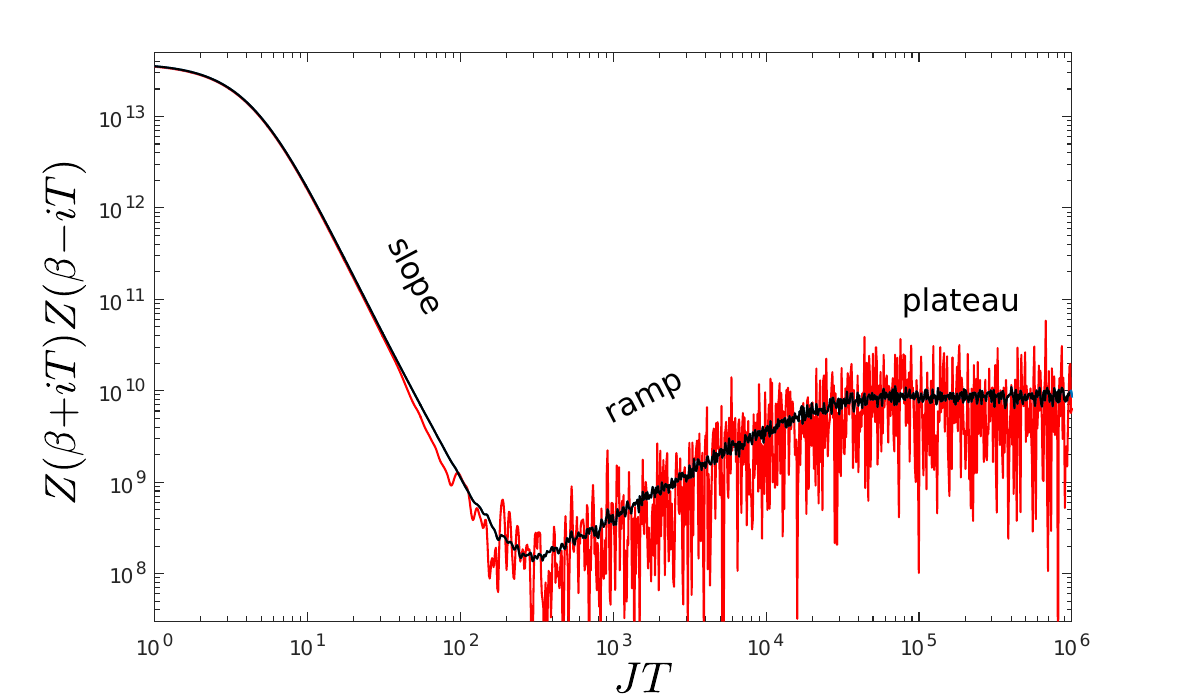}
  \caption{{\small
A log-log plot of the spectral form factor in SYK for $q=4, ~N=34, ~\beta J = 5$ \cite{Cotler:2016fpe}. A single sample (red, erratic) is plotted together with an average of 90 samples (black, smoother). The ramp is approximately linear $\propto T$ in standard variables, not just in the log-log variables.} 
}
  \label{fig:g-SYK}
\end{figure}

The quantity $\langle |Z(\beta+iT)|^2\rangle_J$ for the SYK model was studied numerically in \cite{Cotler:2016fpe}.\footnote{For related work see \cite{ You:2016ldz,Garcia-Garcia:2016mno}.}  (Here and elsewhere in the paper, the angle brackets indicate an average over the ensemble of $J_{a_1...a_q}$ couplings; we will often leave off the $J$ subscript.) We show a plot of this data in figure \ref{fig:g-SYK}.  
The early time decaying ``slope'' region is in some ways the analog of the quasinormal mode behavior for correlators discussed above. In particular, this region of the curve has an explanation involving fluctuations about the naive saddle point in the collective field description or dual AdS${}_2$ gravity theory \cite{Cotler:2016fpe,Stanford:2017thb}.\footnote{For related work on correlation functions see \cite{Bagrets:2016cdf,Bagrets:2017pwq}.} As is evident from the plot, the slope is self-averaging, so that $\langle |Z(\beta+iT)|^2\rangle_J \approx |\langle Z(\beta+iT)\rangle_J|^2$.

The subsequent ``ramp'' and ``plateau'' are not self averaging, and are not consistent with fluctuations about the naive saddle point. However, they are in line with the idea that energy eigenvalues in a chaotic system should follow the statistics of eigenvalues of a random matrix. The spectral form factor is roughly the Fourier transform of the connected two-point correlation function of the eigenvalue density, $\langle \rho(E)\rho(E')\rangle$. In random  matrix theory, a $T$-linear ramp results from a $-1/(E{-}E')^2$ term in this quantity, reflecting long-range repulsion between eigenvalues (see \cite{mehta2004random}, or \cite{Liu:2018hlr} for a recent overview). The plateau arises from a modification of this power-law form once we consider very small energy differences that are of order the typical level spacing $E-E'\sim e^{-S}$. The transition takes place at a time of order the inverse of this spacing, $T\sim e^{S}$.

\subsection{This paper}
Universality of random matrix statistics suggests that a similar ramp and plateau are expected in many physical systems. So from the Hilbert space perspective it is not a surprise that the SYK model or even a black hole should have such behavior. The challenge, motivated by the black hole information problem, is to explain this behavior {\it using the large $N$ collective field variables}, i.e. $G,\Sigma$ for the SYK model or the bulk gravity variables for a black hole. 
In this paper we report some progress on this problem, giving an explanation of the intermediate ramp region in terms of saddle points for $G,\Sigma$, and more speculatively in gravity.   We do not yet have a clear understanding of the plateau,  but will make some preliminary comments in the Discussion.

In {\bf Section \ref{SYK}} we study the SYK model. The $G(t,t'), \Sigma(t,t')$ description is somewhat complicated because of the bilocal nature of the fields, so we warm up in {\bf Section \ref{brownsyk}} by studying a simpler, more random system -- the SYK Brownian circuit.  Here we allow the random couplings $J_{a_1...a_q}$ to vary independently at each time, resulting in local collective fields $G(t), \Sigma(t)$.  The ensemble of time evolution operators $U(T)$ produced by these time dependent Hamiltonians should converge to a uniform Haar measure on a subgroup of unitaries consistent with the time-reversal and fermion parity symmetries. An analog of the averaged spectral form factor is $\langle \Tr [U(T)] \Tr [U(T)]^*\rangle$. At $T=0$ this is equal to $L^2$, where $L = 2^\frac{N}{2}$ is the dimension of the Hilbert space.  At long times it should converge to 
$\int_{\text{Haar}} dU \Tr[ U] \Tr [U]^* $, which is of order one (in fact, precisely two for the relevant unitary ensembles). This is nonvanishing and independent of time, but exponentially smaller than the short time result. One might be tempted to call it a plateau, but we will see that this actually has more in common with the ramp in a time-independent quantum system.

This structure can be explained by saddle points of the $G(t), \Sigma(t)$ system.  Because there are two quantum systems, L and R, often called replicas, there are ``off diagonal'' collective fields $G_{LR}, \Sigma_{LR}$ that couple them.  At short time the $G_{LR} = \Sigma_{LR} = 0$ saddle point with a large negative action dominates.  But its contribution decreases exponentially and eventually allows the contribution from another saddle point, where $G$ and $\Sigma$ assume nonzero constant values, to dominate. This nontrivial saddle point has action equal to zero, and hence gives a contribution of order one at late times, as required.

We begin to study the ramp in {\bf Section \ref{zk}} by repeating the same Brownian circuit $k$ times.   Here $k$ is a discrete analog of SYK time.  The analog of the spectral form factor will now be $\langle \Tr[U(T)^k]\Tr[U(T)^k]^*\rangle$.  The value of this in the standard Haar unitary ensemble is $\text{min}(k,L)$, which is a unitary version of the ramp/plateau structure.  In the $G, \Sigma$ description (suppressing a subtlety to be discussed below) we find $k$ distinct saddle points, each with zero action. This can be explained as follows. The system has a $\mathbb{Z}_k\times \mathbb{Z}_k$ symmetry, an approximation of the $U(1)\times U(1)$ symmetry in the regular SYK case. This is broken to $\mathbb{Z}_k$ by one of the saddles. The $k$ different saddles are simply the orbit of the broken symmetry. 

The total contribution of these saddle points is proportional to $k$, as needed. However, the existence of a plateau here for large enough $k$ demonstrates that there is more to the $G, \Sigma$ description than these saddle points. We make some preliminary remarks about this in Appendix \ref{finiteL}.

Armed with these intuitions we proceed to study regular SYK in {\bf Section \ref{regularsyk}}.  At early times the answer is dominated by independent saddle points of the decoupled L and R systems (given by nonzero $G_{LL}, \Sigma_{LL}, G_{RR}, \Sigma_{RR}$) and the fluctuations around them. For times much larger than $1/J$, these fluctuations are well described by two copies of the Schwarzian theory. The total effect is a large contribution, proportional to $e^{2S_0}$ where $S_0$ is the zero-temperature entropy, but decaying like a power law \cite{Cotler:2016fpe}. This decay is visible in the slope region of figure \ref{fig:g-SYK}.

At late times, the contribution of other saddle points dominate. As in the Brownian SYK case, these have nonvanishing values of  $G_{LR}, \Sigma_{LR}$, correlating the two replicas together. First we discuss the case where $\beta = 0$, where the solutions are approximately as follows. One starts with the correlators in the thermofield double state at an arbitrary auxiliary energy. Then one sums over images to make a function that is antiperiodic in real time. This antiperiodicity is required for the solution to contribute to $Z(iT)Z(-iT)$ (the familiar Euclidean antiperiodicity for $Z(\beta)$ becomes Lorentzian antiperiodicity by time $\pm T$ for the two factors in $Z(iT)Z(-iT)$). Explicit solutions can be found numerically, or analytically at low energies in the conformal limit of SYK. An important point is that the original $\langle Z(iT)Z(-iT)\rangle$ problem has a $U(1) \times U(1)$ symmetry corresponding to independent time translations on the $L$ and $R$ systems. Saddle points with nonzero $G_{LR},\Sigma_{LR}$ spontaneously break this down to the diagonal $U(1)$, analogous to the $\langle \Tr[U(T)^k]\Tr[U(T)^k]^*\rangle$ situation discussed above. We therefore have an exact zero mode, with compact volume proportional to $T$, the size of the periodic circles. So, roughly speaking, there are actually $T$ such saddles.  Each one has action zero, so their sum gives a contribution linear in $T$, the ramp.

For the problem $\langle Z(\beta+iT)Z(\beta-iT)\rangle$ with nonzero $\beta$, there is a complication. In this case we don't quite find a solution to the saddle point equations, due to a pressure towards lower energies. This can be stabilized by taking smoothed microcanonical transforms on both systems: 
\begin{align}\label{microcanonicalTransform}
Y_{E,\DeltaE}(T) &\equiv \int_{\gamma + i \mathbb{R}} d\beta\, e^{\beta E + \beta^2 \DeltaE^2}Z(\beta+iT) \\
|Y_{E,\DeltaE}(T)|^2 &= \int_{\gamma + i \mathbb{R}} d\beta_L\, e^{\beta_L E + \beta_L^2 \DeltaE^2}Z(\beta_L+iT)\int_{\gamma + i \mathbb{R}}d\beta_R \, e^{\beta_R E + \beta_R^2\DeltaE^2}Z(\beta_R-iT).
\end{align}
To study this quantity by saddle points, we allow $\beta_L,\beta_R$ to vary, in addition to the collective fields $G,\Sigma$. Now one finds a saddle point at $\beta_L = \beta_R = 0$ corresponding to the periodically identified thermofield double state described above. Requiring stationarity with respect to variations of $\beta_L,\beta_R$ fixes the energy of the thermofield double state to be $E$. Again, the action is zero and the zero mode described above gives a factor of $T$.

At late times the ramp behavior ends in the plateau.  We do not understand the plateau in the collective description, but we make some preliminary comments in the Discussion.

In {\bf Section \ref{sec:gravity}} we attempt to interpret and generalize these saddle points in gravity. To start, we use that for low energies, the SYK saddle points discussed above can be described in two copies of the Schwarzian theory. So they can be understood as a configuration in the  bulk JT dilaton theory. The configuration corresponds to a Lorentzian wormhole\footnote{Traversable wormhole configurations in SYK with LR coupling have been constructed in \cite{Maldacena:2018lmt}.  Connections to the present work are touched on in sections \ref{nonzerobeta} and \ref{sec:gravity} .   The implications of the existence of wormholes in JT gravity for factorization has been discussed in \cite{Harlow:2018tqv}.   This and related work is commented on in section \ref{singlesample}.}
 connecting the L and R sides with periodic time identification. The family of saddles arises from performing a relative time shift of the L and R boundaries.
Such time shifted wormhole configurations are present in more general gravitational theories. One simply takes a stationary two-sided black hole and periodically identifies in Killing time. This forms a type of Lorentzian double cone. Naively, there is a conical singularity at the horizon due to a fixed point of the identification. We argue that this can be treated with an $i\epsilon$ prescription that follows from (\ref{microcanonicalTransform}) and avoids the singularity.

This configuration may well provide part of the explanation for the universality of random matrix behavior in the boundary dual of these systems. One important issue here is that standard gauge/gravity duals like Super Yang-Mills do not involve averaging over couplings.  In this non-disordered situation, we expect a signal like that in SYK with one fixed realization of the disorder, as in the erratic curve of figure \ref{fig:g-SYK}.  The ramp/plateau structure is clearly visible, but there are fluctuations of size comparable to the signal. The shortest time scale of oscillation is very rapid, of order the inverse width of the energy distribution.  The autocorrelation time is very short as well.  These strong fluctuations pose a challenge, since the geometry just described appears to give a smooth ramp. 

We comment further on this point in the {\bf Discussion}, along with some preliminary discussion of how similar saddle points could contribute to the late-time behavior of correlation functions. We also comment on a possible origin of the plateau.

\section{SYK}\label{SYK}
\begin{figure}[t]
\begin{center}
\includegraphics[width=.8\textwidth]{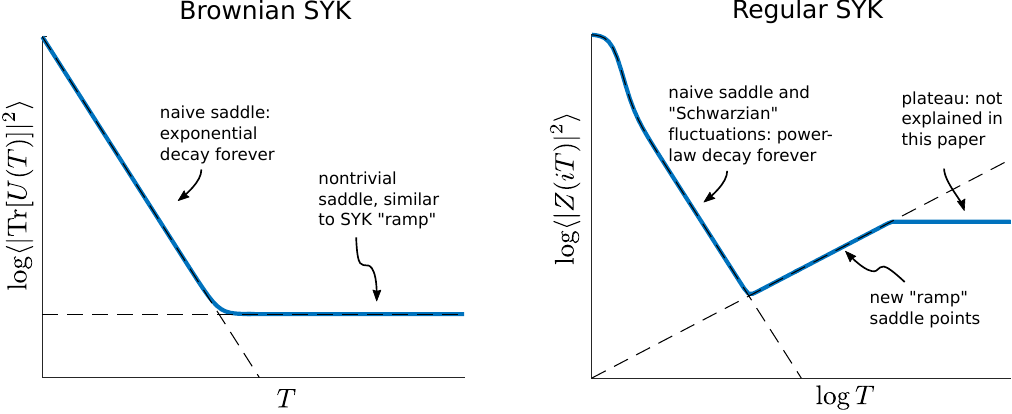}
\caption{{\small In the Brownian model, the $O(1)$ late time value arises from a nontrivial saddle point in the collective field description. In regular SYK, we propose a similar explanation for the ramp. Both plots are simply sketches. (In regular SYK at $\beta = 0$ there are regular oscillations in the slope region. We are sketching the envelope.)}}
\end{center}\label{sketches}
\end{figure}

In this section we discuss saddle points that are important for late-time partition functions in the Brownian SYK model and the regular SYK model. These are saddle points that have nontrivial correlation between the two replicas. In the Brownian model, the saddle point explains the late-time $O(1)$ value of $\langle \Tr[U(T)]\Tr[U(T)]^*\rangle$. In the regular model, we find a family of saddle points that explain a linearly growing ramp in $\langle Z(iT)Z(-iT)\rangle$. The basic features are sketched in figure \ref{sketches}.

For both models, the Hamiltonian is of the form
\be\label{SYKH}
H(t) = i^{\frac{q}{2}}\sum_{a_1<\dots<a_q}J_{a_1\dots a_q}(t)\psi_{a_1}\dots \psi_{a_q}.
\ee
In the regular SYK model, the couplings are assumed to be constant, so that $H$ is time-independent. Ensemble averages are defined by taking the $J_{a_1...a_q}$ variables to be Gaussian distributed with mean zero and variance 
\be
\langle J_{a_1\dots a_q}J_{a_1'\dots a_q'}\rangle = \delta_{a_1a_1'
}\dots \delta_{a_qa_q'}\frac{J^2(q{-}1)!}{N^{q-1}}, \hspace{20pt} \text{(regular SYK)}.
\ee
Here, $J$ sets the energy scale of the system. When we integrate the couplings over such a distribution, we are integrating over variables that are constant in time, which leads to a somewhat inconvenient bilocal action for the fermion variables.

The Brownian SYK model is the opposite extreme, where the couplings are drawn independently at each instant of time. Again, they are Gaussian distributed, with mean zero and variance
\be
\langle J_{a_1\dots a_q}(t)J_{a_1'\dots a_q'}(t')\rangle = \delta_{a_1a_1'
}\dots \delta_{a_qa_q'}\,\delta(t-t')\frac{J(q{-}1)!}{N^{q-1}},\hspace{20pt} \text{(Brownian SYK)}.
\ee
The $\delta(t-t')$ factor means that the couplings are correlated only at the same instant of time. This means that integrating over the couplings leads to a local effective action, which makes the model simpler than regular SYK. It is sometimes useful to imagine replacing this delta function with a sharply peaked but smooth function. This model is an example of a ``Brownian circuit'' studied previously in \cite{Banks:1983by,Lashkari:2011yi}.\footnote{There is also a substantial body of work on the discrete time version of such systems called ``random quantum circuits."  See for example \cite{ Emerson2098,2009CMaPh.291..257H}.}

\subsection{Brownian SYK}\label{brownsyk}
In a time-dependent system like the Brownian SYK model, the time evolution operator is defined as a time-ordered exponential
\be\label{brownianUofT}
U(T) = \mathbf{T}\, e^{-i\int_0^T dt H(t)}.
\ee
For a fixed realization of the time-dependent couplings, we can write a fermion path-integral formula for $\Tr[U(T)]\Tr[U(T)]^* = |\Tr\, U(T)|^2$ as\footnote{The factor of $(-i)^\frac{q}{2}$ in (\ref{yui}) is convenient, and it is intuitive from the perspective of complex conjugation, taking the Grassmann variables to be real. However, replacing $(-i)^\frac{q}{2}$ by $i^\frac{q}{2}$ actually leads to the same answer.}
\begin{align}\label{yui}
|\Tr\,U(T)|^2 = \hspace{-2pt} \int\hspace{-3pt} \mathcal{D}\psi_a^{(L)}\mathcal{D}\psi_a^{(R)}\exp\left\{\hspace{-1pt}i\hspace{-3pt}\int_0^T\hspace{-5pt}dt\left[\frac{i}{2}\psi_a^{(j)}\partial_{t}\psi_a^{(j)} \hspace{-2pt}-\hspace{-2pt}J_{a_1\dots a_q}(t)\left(i^{\frac{q}{2}}\psi_{a_1...a_q}^{(L)} \hspace{-4pt}-(-i)^{\frac{q}{2}} \psi_{a_1...a_q}^{(R)}\right)\right]\right\}.
\end{align}
\vspace{-20pt}

\noindent In this expression, the fermions with replica index $\psi^{(L)}$ compute $\Tr[U(T)]$ and the fermions with replica index $\psi^{(R)}$ compute the complex conjugate $\Tr[U(T)]^*$. The kinetic term involves an implicit sum over $j = L,R$. We used the abbreviation $\psi_{a_1...a_q}\equiv \psi_{a_1}\dots \psi_{a_q}$, and we omitted the time argument for all of the fermion fields, but they are all evaluated at time $t$ inside the integral. In this expression and elsewhere in this paper, the implicit sum over $a_1...a_q$ is over indices in increasing order, $a_1<a_2<...<a_q$.

To take the disorder average, we do the Gaussian integral over all of the separate $J_{a_1\dots a_q}(t)$ variables. This has the effect of squaring the $q$-fermion interaction term in (\ref{yui}). The terms involving all $L$-type or $R$-type fermions can be simplified using $\psi_a\psi_a = \frac{1}{2}$.\footnote{Note that for Grassman integration variables, we usually have $\psi_a(t)\psi_a(t) = 0$, but in this case we can imagine that the disorder averaging is smeared slightly in time, so that the product will involve fermions at slightly different time arguments, and then $\psi_a(t+\epsilon)\psi_a(t) = \frac{1}{2}$.} Anticommuting fermions past each other, it follows that e.g. $\psi^{(L)}_{a_1\dots a_q}\psi^{(L)}_{a_1\dots a_q} = (-1)^\frac{q}{2}/2^q$. We are left with
\begin{align}\label{normal}
\int \mathcal{D}\psi_a^{(L)}\mathcal{D}\psi_a^{(R)}\exp\left\{\hspace{-2pt}-\hspace{-2pt}\int_0^T\hspace{-5pt}dt\left[\frac{1}{2}\psi_a^{(j)}\partial_{t}\psi_a^{(j)} +\frac{J(q{-}1)!}{N^{q-1}}\hspace{-3pt}\sum_{a_1<...<a_q}\hspace{-3pt}\left(\frac{1}{2^q}-\psi^{(L)}_{a_1...a_q}\psi^{(R)}_{a_1...a_q}\right)\right]\right\}.\hspace{-5pt}
\end{align}

We will now make a side comment about the above formula. Since the random couplings were uncorrelated in time, the disorder-averaged theory is local in time. In fact, the above expression can be understood as a normal thermal partition function for a spin system, with {\it real} inverse temperature equal to $T$:
\be\label{spintheory}
\langle |\Tr\,U(T)|^2\rangle = \Tr(e^{-T H_{spin}}), \hspace{20pt} H_{spin} \equiv\frac{J(q{-}1)!}{2^qN^{q-1}}\sum_{a_1<\dots <a_q}\left(1-\sigma^{(z)}_{a_1}\sigma^{(z)}_{a_2}\dots \sigma^{(z)}_{a_q}\right).
\ee
In making this correspondence, we have used that we can represent Majorana fermions in terms of spin variables, and we can arrange that $\psi_a^{(L)}\psi_a^{(R)} = \frac{i}{2}\sigma^{(z)}_a$, where $\sigma^{(z)}_a$ is the Pauli $z$ operator acting at site $a$. For short times the partition function is simply the dimension of the Hilbert space of $N$ spins, $2^N$. For long times the factor $e^{-T H_{spin}}$ becomes a projection operator onto the ground states, times $e^{-TE_{0}}$. A ground state is when all spins are aligned, either up or down, and from the above Hamiltonian we find that the ground state energy is zero. So at large times we find $\langle|\Tr\,U(T)|^2\rangle \rightarrow 2$, where the factor of two is for the two ground states.\footnote{We also see that the approach to this limit is determined by the $N$ lowest lying excited states given by a single flipped spin.   This gives a schematic behavior $2+ N e^{-T}+...$ which determines the time to approach the Haar value $T \sim \log N$.  This is consistent with other determinations of this ``ramp," or ``Thouless" time \cite{Kos:2017zjh,Gharibyan:2018jrp,Bertini:2018wlu}.} 

This is similar to an analogous calculation in the Haar random ensemble on unitary matrices (CUE). A Haar analog of the quantity $\langle |\Tr\,U(T)|^2\rangle$ would be $\int dU \Tr[U]\Tr[U]^* = \int dU \Tr[U\otimes U^*]$. A useful fact is that (see appendix \ref{finiteL} for a more general statement)
\be
\int dU \  U \otimes U^* = |\text{MAX}\rangle \langle\text{MAX}|, \hspace{20pt} |\text{MAX}\rangle  \equiv L^{-\frac{1}{2}}\sum_{i = 1}^L|i\rangle \otimes |i\rangle,
\ee
so $\int dU \Tr[U\otimes U^*]$ reduces to the trace of a projection operator onto an entangled state of the two replicas, similar to what we get in the late-time limit of Brownian SYK. The fact that in Brownian SYK there are two states instead of one is due to the $(-1)^F$ symmetry, see appendix \ref{app:latetimes}. 

\subsubsection{Saddle points for \texorpdfstring{$\langle \Tr[U(T)]\Tr[U(T)]^*\rangle$}{<Tr[U(T)]Tr[U(T)]^*>}}
We now return to the main discussion. Our goal is to find the non-decaying behavior of (\ref{normal}) from the large $N$ action, which is described by variables $G_{LR}(t)$ and $\Sigma_{LR}(t)$ with no fermion flavor indices. The representation of the path integral (\ref{normal}) in terms of these variables is
\begin{align}
\langle |\Tr\,U(T)&|^2\rangle \approx \int \mathcal{D}G_{LR}\mathcal{D}\Sigma_{LR} \exp\left\{-\frac{N}{2}\int_0^T dt \left[\frac{2J}{q}\left(\frac{1}{2^q} - i^qG_{LR}(t)^{q}\right) + \Sigma_{LR}(t)G_{LR}(t)\right]\right\}\notag\\
&\hspace{20pt}\times \int \mathcal{D}\psi_a^{(L)}\mathcal{D}\psi_a^{(R)}\exp\left\{-\frac{1}{2}\int_0^T dt \left[\psi_a^{(A)}\partial_{t}\psi_a^{(A)}   - \psi_a^{(L)}(t)\psi_a^{(R)}(t)\Sigma_{LR}(t)\right]\right\}.\hspace{-20pt}\label{GSigma}
\end{align}
This is a somewhat complicated formula. Let us first understand how it is equivalent to (\ref{normal}). If we integrate $\Sigma_{LR}(t)$ over the appropriate contour, it implements a delta function that sets $G_{LR}(t) = \frac{1}{N}\sum_a\psi_a^{(L)}(t)\psi_a^{(R)}(t)$. With this understanding, the $i^qG_{LR}(t)^q$ term gives the fermion interaction term in (\ref{normal}), with the $i^q = (-1)^{\frac{q(q-1)}{2}}$ factor arising from anticommuting the fermions past each other.\footnote{We have written this expression with a $\,\approx\,$ symbol, because we have approximated the action in a way that is correct at leading order in $N$, but misses terms of order one and smaller in powers of $N$. This form is sufficient for the moment, but see \ref{exactapp} for a precise treatment.}

Let's now proceed naively by looking for a saddle point when we take $\Sigma_{LR}$ and $G_{LR}$ to be constant.\footnote{Although it isn't necessary at the present level of approximation, we show in appendix \ref{exactapp} that the functional integral (\ref{GSigma}) actually localizes to constant $G_{LR},\Sigma_{LR}$ configurations.} Then the fermion determinant on the second line of (\ref{GSigma}) is the partition function of a theory with Hamiltonian $H = -\frac{\Sigma_{LR}}{2}\sum_a \psi^{(L)}_a\psi^{(R)}_a$. Each operator $\psi^{(L)}_a\psi^{(R)}_a$ for fixed $a$ has eigenvalues $\pm i/2$, so the result is $(2\cos\frac{T\Sigma_{LR}}{4})^N$. Combining this with the terms from the first line of (\ref{GSigma}), the total $G_{LR},\Sigma_{LR}$ integrand becomes
\be
\exp\left\{N\left[\log (2\cos\frac{T\Sigma_{LR}}{4}) - \frac{JT}{q2^q} + i^q\frac{JT}{q}G_{LR}^q - \frac{T}{2}\Sigma_{LR} G_{LR} \right]\right\}.
\ee
An obvious saddle point is simply $G_{LR} = \Sigma_{LR} = 0$. The interpretation of this saddle is that the two replicas are not correlated. Note that this is a saddle point for all values of $T$. It predicts simple exponential decay,
\be
\langle |\Tr\,U(T)|^2\rangle \sim 2^N\exp\left[-\frac{JTN}{q 2^q}\right].
\ee
This gives the initial downward slope of figure \ref{sketches}.

However, there are other saddle points and we can ask whether one of them gives the right (non-decaying) behavior at large $T$. For large $T$ we have two simple saddle points
\be\label{nontrivialsaddle}
G_{LR} = \pm\frac{i}{2}, \hspace{20pt} \Sigma_{LR} = \mp\frac{iJ}{2^{q-2}}.
\ee
To check that these are solutions to the saddle point equations, it is useful to note that we can replace $\log 2\cos(T\Sigma_{LR}/4)$ for large values of $T$ by $\pm iT\Sigma_{LR}/4$, depending on the sign of the imaginary part of $\Sigma_{LR}$. An interesting property is that the whole action vanishes for such configurations, and in particular is independent of $T$. So these saddle points predict a non-decaying late time value that is of order one. This is the correct answer, as we found from the spin Hamiltonian discussion. These two saddle points correspond to the two ground states of the effective spin Hamiltonian, in the sense that the saddle point values of $G_{LR}$ are equal to the values of the $\psi_L\psi_R$ correlator in the two ground states.
So in this model, we get the non-decaying late time behavior from a nontrivial saddle point that correlates the two replicas together.

\subsubsection{Saddle points for $ \langle \Tr[U(T)^k]\Tr[U(T)^k]^*\rangle$}\label{zk}
As a generalization of the calculation described above we can consider the same theory, but try to compute $\langle \Tr[U(T)^k]\Tr[U(T)^k]^*\rangle$. This quantity  is the analog of the spectral form factor for unitary groups whose eigenvalues live on the circle. For example, for Haar random unitaries \cite{diaconis2001linear} (see also \cite{pastur2004moments})
\begin{eqnarray}\label{haarkanswer}
\int dU\, \Tr[U^k]\Tr[U^k]^* &= \text{min}(k,L)
  \end{eqnarray}
  where $L$ is the dimension of the matrices. This is the analog of the ramp/plateau structure for random unitary matrices.   

To evaluate this quantity in Brownian SYK  we need to use $2k$ replicas, with fermions $\psi^{(L,s)}_a$ computing the $k$ factors of $U(T)$, and fermions $\psi^{(R,s)}_a$ computing the factors of $U^*(T)$. Here $a = 1,\ldots,N$ is the usual flavor index, and $s = 1,\ldots,k$ is the new replica index.  One can write down a $G,\Sigma$ description for this path integral, where both are now matrices in the $s,s'$ indices.  In particular $G_{ij}^{s s'}(t) = \frac{1}{N}\sum_a\psi_a^{(i,s)}(t) \psi_a^{(j,s')}(t)$ where $i,j$ denote $L$ or $R$, and $\Sigma_{ij}^{s s'}(t)$ is the corresponding Lagrange multiplier.\footnote{Note that $G_{LL}^{ss},  G_{RR}^{ss}$ are equal to $\psi(t+\epsilon)\psi(t) = \frac{1}{2}$.}  The $\psi^{(L,s)}(t)$ form the $k$ segments of the $L$ path integral whose total time is $k T$, and similarly for the $\psi^{(R,s)}(t)$ on the $R$ side.  In order to glue the segments together to form the quantity $\Tr[U(T)^k]$ or $\Tr[U(T)^k]^*$, we need to impose the boundary conditions 
\be\label{bckreplica}
\psi^{(i,s)}(T) = \psi^{(i,s+1)}(0) \hspace{20pt} (s<k), \hspace{50pt} \psi^{(i,k)}(T) = -\psi^{(i,1)}(0).
\ee
The boundary conditions for $G$ follow from this. Note that the problem has a natural $\mathbb{Z}_k \times \mathbb{Z}_k$ symmetry, one factor acting on the $L$ system, one on the $R$.  The $s,s'$ matrix structure forms a discrete analog of the bilocality of the full SYK model.

We will not describe this problem in detail, but as in the case $k = 1$, one can find saddle points that correspond to non-decaying contributions for large $T$. A natural set of such  saddles corresponds to simply pairing up the index $s$ with the index $s'$  according to some permutation belonging to $S_k$, and then setting $G_{LR}^{s s'}, \Sigma_{LR}^{ss'}$ equal to the values in Equation \eqref{nontrivialsaddle} for the paired $s,s'$ values and zero otherwise (all $LL, RR$ quantities for $s\neq s'$ are set to zero as well). For large $T$ this gives a saddle point with zero action, but in order for the configuration to respect the  boundary conditions, the permutation has to be cyclic. There are $k$ cyclic permutations, giving an answer proportional to $k$.(In fact, for each permutation we actually have the choice of setting all the  $G_{LR}^{s s'}, \Sigma_{LR}^{ss'}$  to the same $+$ or $-$ values in \eqref{nontrivialsaddle} so we have $2k$ saddles, giving the expected factor of two for the two decoupled symmetry sectors of the problem, as in the $k=1$ case.)

This is the basic saddle-point origin of the $k$-linear ramp in (\ref{haarkanswer}). Note that these saddles spontaneously break the  $\mathbb{Z}_k \times \mathbb{Z}_k$ symmetry down to a diagonal $\mathbb{Z}_k$, and the $k$ distinct saddles form an orbit of the broken symmetry.  This symmetry pattern provides a clue about the type of saddle to look for in the regular SYK model, which we will turn to in the next section.\footnote{The relationship of the ramp to such cyclic permutations occurs in many contexts.  The symmetry breaking pattern appears in the supersymmetric sigma model approach \cite{Efetov:1997fw}. The diagrams in GUE giving the ramp are cyclically permuted ladders \cite{Brezin:1993qg}. In CUE the factor of $k$ arises from cyclically permuted identifications of $U$ with $U^*$, much as here. Cyclic permutations have also played a central role in recent work deriving the ramp for Floquet many-body system \cite{Kos:2017zjh,Chan:2017kzq,Chan:2018dzt,Bertini:2018wlu} .} 

In addition to these saddle points, when $q = 2$ (mod 4), there are additional saddles described in appendix \ref{app:Nmod8}. After including this detail, the above saddle points seem consistent with the ensembles of unitary matrices that we expect the Brownian circuit to converge to at late times, see appendix \ref{app:latetimes}. However, there is more to this model, because the behavior for such ensembles changes qualitatively $k > 2^{N/2 -1}$ or $k > 2^{N/2 -2}$, exhibiting a plateau phenomenon similar to the one exhibited in(\ref{haarkanswer}). It would be desirable to have an understanding of this in terms of the $G,\Sigma$ path integral. We make some preliminary remarks about the mechanisms involved in Appendix \ref{finiteL}.

\subsection{Regular SYK}\label{regularsyk}
In the regular SYK model we will find an analog of the nontrivial Brownian SYK saddle points discussed in the previous section. This analog consists of a family of saddle points that appear to explain a linear ramp in $\langle |Z(iT)|^2\rangle$, but not the plateau. As we will see, there is a complication when $\beta$ is nonzero that leads us to consider $\langle |Y_{E,\DeltaE}(T)|^2\rangle$ instead of $\langle |Z(\beta+iT)|^2\rangle$. We will start with the simpler $\beta = 0$ (infinite temperature) case where this doesn't arise.

\subsubsection{The case where $\beta = 0$}\label{secbeta0}
We would like to study the quantity $Z(iT)Z(-iT)$. One can write a path-integral expression for this as in (\ref{yui}), but with constant couplings. Doing the disorder average, we find a bilocal action involving an integral over two times. This action can be represented as an integral over collective fields as in (\ref{GSigma}), except that we now integrate over more variables. Instead of the variables $G_{LR}(t),\Sigma_{LR}(t)$, we now have matrices $G_{ij}(t,t'), \Sigma_{ij}(t,t')$ where $i,j \in \{L,R\}$:
\begin{align}\label{GSigmaSYK}
&\langle Z(iT)Z(-iT)\rangle = \int \mathcal{D}G\mathcal{D}\Sigma e^{-N\, I[G,\Sigma]}\notag\\
&I[G,\Sigma] = -\text{log Pf}\left(\delta_{ij}\partial_t - \Sigma_{ij}\right) + \frac{1}{2}\int_0^T\int_0^T dt dt'\left( \Sigma_{ij}G_{ij} -\frac{J^2}{q}s_{ij}G_{ij}^{q} \right)\\
&s_{LL}= s_{RR} = -1, \hspace{20pt} s_{LR} = s_{RL} = i^q = (-1)^\frac{q}{2}.\notag
\end{align}
Here we left off the time arguments, but in all cases $G$ and $\Sigma$ depend on a pair of times $t,t'$. We wrote the analog of the fermion path integral from (\ref{GSigma}) in a compact way as a Pfaffian.

At large $N$, the path integral in (\ref{GSigmaSYK}) is semiclassical, dominated by saddle points and small fluctuations around them. It isn't obvious that the ramp should be visible in a semiclassical approximation to the $G,\Sigma$ integral, but we will in fact find a reasonable-seeming ramp within this approximation. To get the saddle point equations, one can vary the action with respect to $G$ and $\Sigma$. In order to simplify the resulting equations, we will make an ansatz that $G,\Sigma$ are only functions of the difference of times, e.g.
\be
G_{ij}(t,t') = \hat{G}_{ij}(t-t').
\ee
In what follows we will omit the hat, using the argument of $G$ to distinguish. It then becomes convenient to decompose the functions into Fourier components. In order to compute $Z(iT)$ or $Z(-iT)$, we are working on a real-time circle of length $T$. The functions $G,\Sigma$ inherit the antiperiodicity around this circle associated with the trace boundary conditions for the original SYK fermions. So $G,\Sigma$ should be antiperiodic with respect to either of their time arguments. This means that the frequencies that appear are fermionic (half-integer) Matsubara frequencies, e.g.
\be
G_{ij}(\omega_n) = \int_0^T dt e^{i\omega_n t}G_{ij}(t), \hspace{20pt} \omega_n = \frac{2\pi (n+\frac{1}{2})}{T}.
\ee
The saddle point equations can then be written as
\begin{align}\label{saddlePt}
\left(\begin{array}{cc}G_{LL}(\omega_n) & G_{LR}(\omega_n)\\ G_{RL}(\omega_n) & G_{RR}(\omega_n)\end{array}\right) &= -\left(\begin{array}{cc}i\omega_n + \Sigma_{LL}(\omega_n) & \Sigma_{LR}(\omega_n)\\ \Sigma_{RL}(\omega_n) & i\omega_n + \Sigma_{RR}(\omega_n)\end{array}\right)^{-1}\\
\Sigma_{ij}(t) &= s_{ij}J^2G_{ij}^{q-1}(t).\notag\end{align}
We can reduce the number of variables somewhat by using that $G_{ij}(t,t') = -G_{ji}(t',t)$, which implies that $G_{ij}(t) = -G_{ji}(-t)$ and $G_{ij}(\omega_n) = -G_{ji}(-\omega_n)$ and similarly for $\Sigma$. For example, this allows us to write $G_{RL}$ and $\Sigma_{RL}$ in terms of $G_{LR}$ and $\Sigma_{LR}$.

If we assume $G_{LR} = \Sigma_{LR} = 0$, then we find the decoupled equations for the $L$ and $R$ ~SYK systems.  The solutions to these equations, and the diagonal fluctuations about them, are simply computing $\langle Z(iT)\rangle\langle Z(-iT)\rangle$ which tends to zero at large time $T$. This situation is similar to the trivial $G_{LR} = 0$ saddle that we found in the Brownian SYK model. Guided by the Brownian SYK example, we would like to find nontrivial saddles where $G_{LR}$ is nonzero.

\begin{figure}[t]
\begin{center}
\includegraphics[width=.5\textwidth]{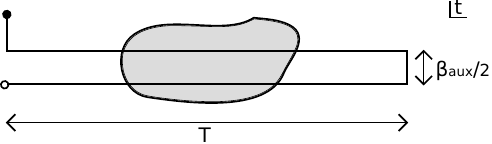}
\caption{\small{We show a path-integral repesentation of the quantity $\text{Tr}(e^{-\frac{\beta_{\text{aux}}}{2}H}e^{-iHT}e^{-\frac{\beta_{\text{aux}}}{2}H}e^{iHT})$. The open and filled circles should be identified. In SYK one could study this quantity by path $G,\Sigma$ path integral. The leading saddle point is simply the analytic continuation to this contour of the standard thermal saddle point for $Z(\beta_{\text{aux}})$. The idea is that the equations for $G,\Sigma$ in the shaded region are similar to our equations, with $L$ and $R$ referring two the two sides of the contour, (and with a factor of $i^q$ that arises from a different convention for $G_{LR}$).}}\label{figtwocontours}
\end{center}
\end{figure}

We will now give some motivation for the existence of nontrivial solutions, before discussing them explicitly. Let's temporarily forget about the $Z(iT)Z(-iT)$ quantity and consider instead $\text{Tr}(e^{-\frac{\beta_{\text{aux}}}{2}H}e^{-iHT}e^{-\frac{\beta_{\text{aux}}}{2}H}e^{iHT})$, where we introduced an arbitrary parameter $\beta_{\text{aux}}$. Of course, the real-time evolution trivially cancels in this quantity, which reduces to $Z(\beta_{\text{aux}})$. Still, we can imagine computing it by an elaborate path integral that represents each of the four factors explicitly, see figure \ref{figtwocontours}. The saddle point configuration for $G,\Sigma$ on this contour will simply be the analytic continuation of the thermal correlation functions at inverse temperature $\beta_{\text{aux}}$. Now, the basic idea is that the saddle point equations for $G,\Sigma$ in the long Lorentzian parts of the contour are approximately the same as the equations for our $Z(iT)Z(-iT)$ problem, after identifying the two folds of the contour with the $L$ and $R$ systems. The only difference between the problems is at the ends: in the $Z(iT)Z(-iT)$ case we separately periodically identify both the $L$ and $R$ contours, and in the case in figure \ref{figtwocontours} we glue the contours together after Euclidean evolution by $\beta_{\text{aux}}/2$. \footnote{One might be suspicious of this logic, because the $G,\Sigma$ equations are nonlocal in time, so the difference at the ends might be significant. However, the nonlocality is controlled by the bilocality of $G(t,t')$ and $\Sigma(t,t')$, which are exponentially decaying in the time separation, so if $T$ is large we expect the equations to be local enough to make the argument.}

The above motivates us to construct approximate solutions to (\ref{saddlePt}) using the analytically continued thermal correlators at an arbitrary inverse temperature $\beta_{\text{aux}}$. More precisely, we would like to take as a trial solution the correlators in the thermofield double state for the $L$ and $R$ systems defined for $t>0$ as
\be
G^{(\beta_{\text{aux}})}_{ij}(t) = \langle TFD(\beta_{\text{aux}})|\psi^{(i)}(t)\psi^{(j)}(0)|TFD(\beta_{\text{aux}})\rangle, \hspace{20pt} i,j \in \{L,R\}.
\ee
However, these correlators do not satisfy the correct boundary conditions; they are exponentially decaying for large $|t|$ and in particular are not antiperiodic with period $T$. We can fix this by summing over images, or more simply taking
\be\label{guess}
G_{ij}(t) = G^{(\beta_{\text{aux}})}_{ij}(t) - G^{(\beta_{\text{aux}})}_{ij}(t-T), \hspace{20pt} 0<t<T,
\ee
and antiperiodically extending outside this range. We do not expect this to be an exact solution, but if $T$ is large we expect there to be a true solution very close by.

We can make this argument a little more precise as follows. Let's imagine trying to verify that (\ref{guess}) is a solution. We will need to compute the Fourier transform, which we can treat as
\begin{align}
G_{ij}(\omega_n) &= \int_0^T dt e^{i\omega_n t}G_{ij}(t) = \int_{-\frac{T}{2}}^{\frac{T}{2}}dt e^{i\omega_n t} G_{ij}(t) \approx \int_{-\frac{T}{2}}^{\frac{T}{2}}dt e^{i\omega_n t}G^{(\beta_{\text{aux}})}_{ij}(t)\\
&\approx \int_{-\infty}^\infty dt e^{i\omega_n t}G_{ij}^{(\beta_{\text{aux}})}(t).\label{logicin}
\end{align}
In the last step of the first line, we used that for $0<t<T/2$, the second term $G^{(\beta_{\text{aux}})}_{ij}(t-T)$ in (\ref{guess}) is much smaller than the first term $G^{(\beta_{\text{aux}})}_{ij}(t)$. In going to the second line, we used that for large $T$, the error we make in extending the region of integration is exponentially small in $T$. Now, the point is that $G_{ij}^{(\beta_{\text{aux}})}$ is actually a solution on the infinite line, so after making these approximations (and similar approximations for $\Sigma$) we will find that the equations are satisfied.

\begin{figure}[t]
\begin{center}
\includegraphics[width=.32\textwidth]{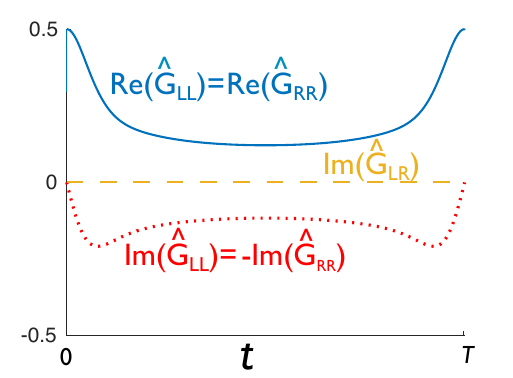}
\includegraphics[width=.32\textwidth]{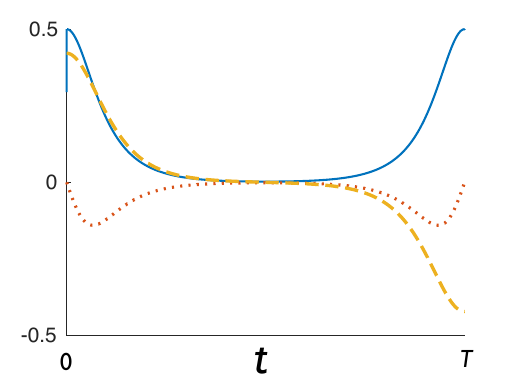}
\includegraphics[width=.32\textwidth]{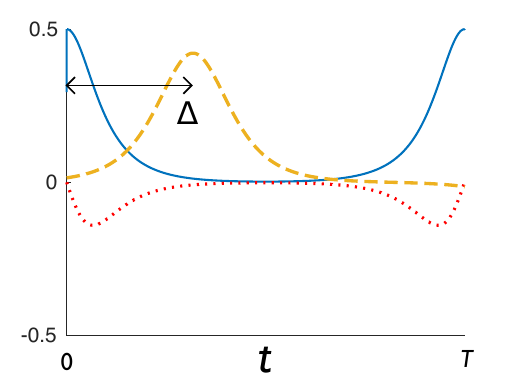}
\caption{\small{Example numerical solutions to (\ref{saddlePt}). {\bf Left:} the disconnected saddle point with $G_{LR} = 0$. Solid blue is $\text{Re}(G_{LL}) = \text{Re}(G_{RR})$, dotted red is $\text{Im}(G_{LL}) = -\text{Im}(G_{RR})$, and dashed yellow is $\text{Im}(G_{LR})$. {\bf Middle:} a connected saddle point. {\bf Right:} the same solution but with nonzero $\Delta$. For the connected solutions, increasing $\beta_{\text{aux}}$ would broaden the features and increase the magnitude of the imaginary part of $G_{LL}$. Increasing $T$ would extend the middle part of the plots where the solutions are very small.}}\label{fignumericalplots}
\end{center}
\end{figure}

This argument shows that our configuration {\it almost} satisfies the saddle point equations, but one would still like to show that there is a nearby {\it exact} solution. This can be done by solving the equations by numerical iteration, with (\ref{guess}) as a starting point. In figure \ref{fignumericalplots} we show plots of numerical solutions. Although we do not plot (\ref{guess}) for comparison, they would be indistinguishable by eye already for these modest values of $T$.
There are two crucial features of these solutions that we need to comment on. First, the solutions have a saddle point action that is very close to zero. This is easy to check numerically. Intuitively, it is reasonable that the action should be independent of $T$ for large $T$, since the action of the long Lorentzian parts of the contour in figure \ref{figtwocontours} must be zero in order for that quantity to be independent of time. However, to get the right constant value for the action requires a further argument that we give in appendix \ref{actionapp} where we show that the classical action is zero up to exponentially small (in $T$) corrections.\footnote{This implies that the weight of this saddle point is schematically 
$\exp(N(0+ e^{-T/\beta_{\text{aux}}}))$.   So the time  at which its contribution becomes close to the random matrix value is of order  $t_{\text{ramp}} \sim \beta_{\text{aux}}\log N$.  This is the  ``ramp" or Thouless time  (defined for example in \cite{Gharibyan:2018jrp} ) of the SYK model.} Note that for early times, the standard uncorrelated solutions have a negative action, so the correlated configuration we are describing here is exponentially subleading (in $N$) at early times.

The second crucial feature is that these solutions are actually part of a two-parameter family. One parameter is the arbitrary $\beta_{\text{aux}}$ that we have discussed above. The second parameter can be understood as follows. The action (\ref{GSigmaSYK}) has independent time translation symmetries on the $L$ and $R$ systems. The solution we are considering spontaneously breaks this symmetry by correlating the fermions on the $L$ system with fermions on the $R$ system at the same time. What remains is only a diagonal time translation symmetry. By acting on this solution with the spontaneously broken symmetry generator, which shifts the time in opposite directions on the $L$ and $R$ systems, we generate new solutions, where the fermions on the $L$ system are correlated with fermions on the $R$ system at a different time. Concretely, we generate new solutions by leaving $G_{LL},G_{RR}$ the same, but sending $G_{LR}(t)\rightarrow G_{LR}(t-\Deltat)$. The quantity $\Deltat$ is the second parameter in our two-parameter family of solutions. Since the functions are antiperiodic with period $T$, we find that $\Deltat$ is valued on a circle of circumference $2T$, and integrating over this zero mode gives us a factor of $2T$.\footnote{At first it seems that there could be a third zero mode, which corresponds to using the $G,\Sigma$ configuration for the quantity $\text{Tr}(e^{-\frac{\beta_{\text{aux}}+x}{2}H}e^{-iHT}e^{-\frac{\beta_{\text{aux}}-x}{2}H}e^{iHT})$ with $x$ a new variable. In fact, $x$ is just the imaginary part of $\Deltat$, so it is not a new mode. In the path integral we should do a contour integral over $\Deltat$, and we are assuming the defining contour is in the real (periodic) direction. This is consistent with the contour for the nonzero modes, see appendix \ref{app:oneloop}.}

The contribution of these saddle points is then of the form
\be\label{ourramp}
\langle |Z(iT)|^2\rangle \supset \int_0^\infty d\beta_{\text{aux}}\mu(\beta_{\text{aux}})\int_0^{2T} d\Deltat = (const.) \,T.
\ee
Here the measure factor $\mu(\beta_{\text{aux}})$ comes from a fluctuation integral around the saddle point. In appendix \ref{app:oneloop} we evaluate the one-loop determinant for $TJ\gg 1$, and we find that measure is such that $\int d\beta_{\text{aux}}\mu(\beta_{\text{aux}})$ becomes an integral over energies $E_{\text{aux}}$ with a flat measure $\frac{dE_{\text{aux}}}{2\pi}$. This is the final answer for the case where $q \equiv 2$ (mod 4):
\be\label{q6}
\langle |Z(iT)|^2\rangle \supset 2T\int\frac{dE_{\text{aux}}}{2\pi}, \hspace{30pt} \text{when $q \equiv 2$ (mod 4)}.
\ee
In the case $q = 4$ there is a time-reversing $\mathbb{Z}_2$ symmetry of the saddle point equations that leads to another set of solutions and therefore a further factor of two:
\be\label{q4}
\langle |Z(iT)|^2\rangle \supset 4T\int\frac{dE_{\text{aux}}}{2\pi}, \hspace{30pt} \text{when $q \equiv 0$ (mod 4)}.
\ee
The integral over energy gives $E_{max} - E_{min} = 2E_{max}$ which is proportional to $NJ$, with a coefficient that must be determined by numerical solution of the standard Euclidean Schwinger-Dyson equations (in the case $q = 4$, $E_{max} \approx 0.0406 N J$).

These expressions match the expectations from random matrix theory for the ramp, including the precise numerical factors. Checking this is a little bit nontrivial, since the random matrix ensembles corresponding to the discrete symmetries of SYK (time reversal and fermion parity) depend on both $N$ and $q$. It turns out that the $N$ dependence cancels out in the coefficient of the ramp, as explained in appendix \ref{app:rmtExpectations}.\footnote{We are grateful to Guy Gur-Ari for raising the question of the $N$ dependence, and to David Gross for motivating us to check the numerical coefficient.}

Let us briefly summarize: because of the integral over the zero mode $\Deltat$, the contribution of these saddles is proportional to $T$. In addition, since the action is zero, there is no exponential-in-$N$ prefactor. This matches our rough expectations for the ramp in $\langle |Z(iT)|^2\rangle$ based on random matrix theory. The one-loop determinant precisely matches the expected numerical coefficient.

\subsubsection{Nonzero $\beta$ and $|Y_{E,\DeltaE}(T)|^2$}\label{nonzerobeta}

Now we turn to the problem $\langle |Z(\beta+iT)|^2\rangle$ with $\beta$ nonzero. There is a subtlety in this case, and to explain it we will first review expectations for $\langle |Z(\beta+iT)|^2\rangle$ based on random matrix theory. For simplicity, let's consider the case where the eigenvalues of the Hamiltonian have GUE statistics. In this case (see e.g.~\cite{Cotler:2016fpe} Eq.(42)) we expect to find schematically
\be\label{rmtramp}
\langle |Z(\beta+iT)|^2\rangle \propto \int dE_{\text{aux}}\, \text{min}\left(\tfrac{T}{2\pi},e^{S(E_{\text{aux}})}\right)e^{-2\beta E_{\text{aux}}}.
\ee
The connected saddle points we discussed in the previous section seem to be adequate for explaining the ramp, but not the plateau. So, at best we could hope to find an expression 
\be\label{integrand}
\langle |Z(\beta+iT)|^2\rangle \propto \int dE_{\text{aux}}e^{-2\beta E_{\text{aux}}}\int_0^{2T} d\Deltat .
\ee
Notice that there is now a pressure on $E_{\text{aux}}$ (or the corresponding $\beta_{\text{aux}}$) from the $e^{-2\beta E_{\text{aux}}}$ factor. What this means is that $\beta_{\text{aux}}$ or equivalently $E_{\text{aux}}$ no longer represents a flat direction in the action, instead it represents a direction with a nonzero slope $2\beta$. As a result, we should not expect to find genuine saddle points when $\beta$ is nonzero. This is consistent with numerics: as one iterates the Schwinger-Dyson equations, the solution almost converges to a configuration similar to the $\beta = 0$ solutions, but as the iterations proceed the effective $\beta_{\text{aux}}$ increases slowly but steadily, and we do not find an actual solution.

To get a quantity that can be studied by honest saddle points, one can consider a microcanonical version of the spectral form factor, 
\be\label{Y}
|Y_{E,\DeltaE}(T)|^2 = \int_{\gamma + i \mathbb{R}} d\beta_L\, e^{\beta_L E + \beta_L^2 \DeltaE^2}Z(\beta_L+iT)\int_{\gamma + i \mathbb{R}}d\beta_R \, e^{\beta_R E + \beta_R^2\DeltaE^2}Z(\beta_R-iT).
\ee
This quantity is sensitive to the contributions where both the $L$ and $R$ energies are within roughly $\DeltaE$ of the energy $E$. This should provide a stabilization that gives a saddle point for $E_{\text{aux}}$ or equivalently $\beta_{\text{aux}}$. 

Let's describe how this works in more detail. After the disorder average, $\langle |Y_{E,\DeltaE}(T)|^2\rangle$ is described by a $G,\Sigma$ action. The only difference is that we will be looking for stationary points with respect to all variables including $\beta_L,\beta_R$. To analyze this, a convenient trick is to view the partition function $Z(\beta+iT)$ as simply $Z(iT)$ for a theory where we multiply the Hamiltonian by $(1 - \frac{i\beta}{T})$. In the SYK model, multiplying the Hamiltonian by something can be accomplished by rescaling $J$. So we define
\be
J_L = (1 - \frac{i\beta_L}{T})J,\hspace{20pt}J_R = (1 + \frac{i\beta_R}{T})J.
\ee
The full action for studying $\langle |Y_{E,\DeltaE}(T)|^2\rangle$ can then be written as
\be
-\left[ (\beta_L+\beta_R) E + (\beta_L^2+\beta_R^2)\DeltaE^2\right]-N\text{log Pf}\left(\delta_{ij}\partial_t \hspace{-2pt}-\hspace{-2pt}\Sigma_{ij}\right) + \frac{N}{2}\int_0^T\int_0^T dt dt'\left[ \Sigma_{ij}G_{ij} - \frac{J_iJ_j}{q}s_{ij}G_{ij}^q\right].
\ee

\vspace{-10pt}
\noindent where as before $s_{LL} = s_{RR} = -1$ and $s_{LR} = s_{RL} = (-1)^\frac{q}{2}$.

The equations we get by varying with respect to $G,\Sigma$ are the same as (\ref{saddlePt}), with $J^2 \rightarrow J_iJ_j$. In addition there are two equations that one gets by varying with respect to $\beta_L,\beta_R$. Because of the $\beta_{\text{aux}}$ instability, we do not expect solutions of the $G,\Sigma$ equations when $\beta_L,\beta_R$ are nonzero.\footnote{A subtlety: there can be solutions when $\beta_L = -\beta_R$, which is a flat direction degenerate with changing $T$. This component is set to zero by requiring stationarity of the $\DeltaE^2$ term.} So the $G,\Sigma$ equations effectively set $\beta_L,\beta_R$ to zero. Naively, this lands us back on the $\langle |Z(iT)|^2\rangle$ problem, for which we argued that there is a two-parameter family of solutions, labeled by $\beta_{\text{aux}}$ and $\Deltat$. However, we still have to impose the equations that we get by varying with respect to $\beta_L,\beta_R$. These give (after setting $\beta_L = \beta_R = 0$)
\be\label{energyconstraint}
E = -\frac{iJ^2N}{qT}\int_0^T\int_0^T dt dt'\left[G_{LL}^q - i^qG_{LR}^q\right] = \frac{iJ^2N}{qT}\int_0^T\int_0^T dt dt'\left[G_{RR}^q - i^qG_{LR}^q\right].
\ee
Within the two-parameter family just mentioned, the middle and rightmost expressions in (\ref{energyconstraint}) are equal to each other, and are a function of $\beta_{\text{aux}}$. Solving (\ref{energyconstraint}) pins down the value of $\beta_{\text{aux}}$ in terms of the argument $E$ of $Y_{E,\DeltaE}(T)$. In fact, the equation has a simple interpretation. Using the saddle point equations (\ref{saddlePt}) and arguing as in (5.69) of \cite{Maldacena:2018lmt}, one can show that the quantity on the RHS is simply the (equal) energies of the $L$ and $R$ systems. The parameter $\Deltat$ is not fixed, however, so integrating over this zero mode, we continue to find an answer proportional to $T$.

The following comment is an aside. The $\beta_{\text{aux}}$ instability has a connection to the recent work of Maldacena and Qi \cite{Maldacena:2018lmt}. We can imagine setting $T = 0$ and studying the quantity $\langle Z(\beta)^2\rangle$ by saddle point. In the original SYK theory, one does not find a connected saddle point that correlates the two theories. However, if the two are coupled, then \cite{Maldacena:2018lmt} shows that a connected solution exists. We can follow a similar strategy here with nonzero $T$. By coupling the $L$ and $R$ theories, connected solutions for $\langle |Z(\beta+iT)|^2\rangle$ can be found. The $L$-$R$ coupling stabilizes $\beta_{\text{aux}}$. In the case of \cite{Maldacena:2018lmt}, it was natural to consider a local-in-time coupling. However, this breaks the independent time translation symmetries and removes the possibility of a ramp, so for our purposes it is more convenient to add a bilocal $L$-$R$ term to the action that is integrated separately over the $L$ and $R$ times. A simple example is to add a multiple of the $G_{LR}^q$ term that is already present in the action, modifying the coefficient $s_{LR} = s_{RL} \rightarrow  (-1)^\frac{q}{2}(1+\lambda)$. One can then find numerical solutions for nonzero $\beta$.

Of course, these are solutions to a different problem, but we can understand them in the sense of constrained instantons \cite{Affleck:1980mp}. When $\beta$ is nonzero, there are ``almost-solutions'' to the original problem, namely configurations of $G,\Sigma$ that are stationary with respect to all but one direction in function space, corresponding to $\beta_{\text{aux}}$. This situation is similar in some respects to that of instantons in massive $\phi^4$ theory, and a useful method in that context is to artificially change the problem by adding a term that stabilizes the one unstable direction \cite{Affleck:1980mp} and reveals the almost-solutions. The extra $\lambda G_{LR}^q$ term can be interpreted this way.

\subsubsection{Further details in the conformal limit}\label{conformalDetails}
It is possible to understand the solutions more concretely in the analytically tractable limit of large $\beta_{\text{aux}}$. This is the conformal limit of SYK, where we can drop the $i\omega_n$ terms in the saddle point equations (\ref{saddlePt}). A small subtlety is that in this limit, the pressure on $\beta_{\text{aux}}$ becomes weak, and one can actually find solutions to the saddle point equations for $\langle Z(\beta+iT)Z(\beta-iT)\rangle$ with nonzero $\beta$. The solutions are of the approximate form (\ref{guess}), with
\begin{align}
G_{LL}^{(\beta_{\text{aux}})}(t) &= \frac{\big(\frac{\beta-iT}{\beta+iT}\big)^{\frac{1}{q}}\,b}{\big|\frac{\beta_{\text{aux}}}{\pi}\sinh(\frac{\pi t}{\beta_{\text{aux}}}\sqrt{1+\frac{\beta^2}{T^2}})\big|^{\frac{2}{q}}}\text{sgn}(t)\label{GLL}\\
G_{LR}^{(\beta_{\text{aux}})}(t) &= \frac{ib}{\big(\frac{\beta_{\text{aux}}}{\pi}\cosh(\frac{\pi (t-\Deltat)}{\beta_{\text{aux}}}\sqrt{1+\frac{\beta^2}{T^2}})\big)^{\frac{2}{q}}}\label{GLR}
\end{align}
where $J^2 b^q\pi = (\frac{1}{2}-\frac{1}{q})\tan\frac{\pi}{q}$. The missing component $G_{RR}$ is given by the complex conjugate of $G_{LL}$. Note that when $\beta/T = 0$, we simply get the correlation functions in the thermofield double state at inverse temperature $\beta_{\text{aux}}$.

We would now like to go beyond the conformal limit slightly, to see the small pressure on $\beta_{\text{aux}}$ and understand how it can be stabilized. For the usual SYK model in Euclidean signature, the leading correction to the conformal limit comes from the Schwarzian action,
\be\label{schorig}
I_{Sch} = -\frac{\alpha_S}{\mathcal{J}}\int d\tau\, \text{Sch}(f,\tau), \hspace{20pt} \text{Sch}(f,x) \equiv \frac{f'''(x)}{f'(x)} - \frac{3}{2}\left(\frac{f''(x)}{f'(x)}\right)^2,
\ee
where the path integral weighting is by $e^{-N I_{Sch}}$, and $\mathcal{J} = \sqrt{q}J/2^{\tfrac{q-1}{2}}$. The variable $f(\tau)$ has the interpretation of a reparametrization of time, and it is related to the $G$ configuration by
\be\label{corr}
G(\tau_1,\tau_2) = b \left(\frac{f'(\tau_1)f'(\tau_2)}{(f(\tau_1)-f(\tau_2))^2}\right)^\frac{1}{q}.
\ee
Concretely, the leading non-conformal correction to the action for a $G$ configuration of the form (\ref{corr}) is the action (\ref{schorig}).

In the present case, we expect two copies of the Schwarzian degree of freedom, for the $L$ and $R$ systems. On the $L$ system the time $t$ is related to Euclidean time by $\tau = (\frac{\beta}{T}+i)t$ and on the $R$ system the two are related by $\tau = (\frac{\beta}{T}-i)t$. Adding two copies of (\ref{schorig}) together and changing variables from $\tau$ to $t$ according to this rule, we arrive at the action
\be\label{fullaction}
I_{Sch} = -\frac{\alpha_S}{\mathcal{J}(\frac{\beta}{T}+i)}\int_0^T dt\,\text{Sch}(f_L,t) -\frac{\alpha_S}{\mathcal{J}(\frac{\beta}{T}-i)}\int_0^T dt\,\text{Sch}(f_R,t)
\ee
The quantities $f_L(t)$ and $f_R(t)$ are related to the correlators (up to image terms that make the answer antiperiodic) by a generalization of (\ref{corr}) where we write the $\partial_\tau$ derivatives in terms of $\partial_t$ on the two sides at the cost of factors of $\frac{\beta}{T}\pm i$, and introduce a phase for the $G_{LR}$ correlator:
\be
G_{LL}(t_1,t_2) = b \left(\frac{\frac{1}{(\frac{\beta}{T}-i)^2}f_L'(t_1)f_L'(t_2)}{(f_L(t_1)-f_L(t_2))^2}\right)^{\frac{1}{q}}, \hspace{15pt} G_{LR}(t_1,t_2) = ib \left(\frac{\frac{-1}{\frac{\beta^2}{T^2}+1}f_L'(t_1)f_R'(t_2)}{(f_L(t_1)-f_R(t_2))^2}\right)^{\frac{1}{q}}
\ee
Our configurations (\ref{GLL}) and (\ref{GLR}) correspond to
\be
f_{L} = \tanh\big(\frac{\pi t}{\beta_{\text{aux}}}\sqrt{1+\frac{\beta^2}{T^2}}\big), \hspace{20pt} f_R = \frac{1}{\tanh\big(\frac{\pi(t+\Deltat)}{\beta_{\text{aux}}}\sqrt{1+\frac{\beta^2}{T^2}}\big)}.
\ee
The full action (\ref{fullaction}) for this configuration can be worked out by plugging in. One finds
\be\label{sch}
I_{Sch} = \frac{\alpha_S}{\mathcal{J}}\frac{4\pi^2\beta}{\beta_{\text{aux}}^2}.
\ee
Note that this is independent of $T$, due to a cancellation between the two terms in (\ref{fullaction}). We see that indeed there is a pressure from $e^{-N I_{Sch}}$ on the $\beta_{\text{aux}}$ parameter, pushing towards larger values of $\beta_{\text{aux}}$. In fact, this is just the factor $e^{-2\beta E(\beta_{\text{aux}})}$ anticipated in (\ref{integrand}). The pressure on $\beta_{\text{aux}}$ is small at large $J\beta_{\text{aux}}$, however, and the conformal limit of the saddle point equations misses it altogether, which makes the solutions (\ref{GLL}) and (\ref{GLR}) possible.

However, for finite $J\beta_{\text{aux}}$ there is a still a small pressure, and we should see how it gets stabilized in the microcanonical $\langle |Y_{E,\DeltaE}(T)|^2\rangle$ quantity. To compute this, we study $Z(\beta_L+iT)Z(\beta_R-iT)$ and integrate over the $\beta_{L,R}$ parameters with an appropriate weighting. This problem can be obtained from $Z(\beta+iT)Z(\beta-iT)$ by taking $\beta \rightarrow \frac{\beta_L+\beta_R}{2}$ and $T\rightarrow T -\frac{\beta_L-\beta_R}{2i}$. Adding together the Schwarzian action (\ref{sch}) and the microcanonical weighting factors from (\ref{Y}), we find the final integral over $\beta_{L,R}$ and $\beta_{\text{aux}}$ as
\be
\langle|Y_{E,\DeltaE}(T)|^2\rangle \sim T \int\hspace{-2pt} d\beta_{\text{aux}} d\beta_L d\beta_R \exp\hspace{-2pt}\left[(\beta_L\hspace{-2pt}+\hspace{-2pt}\beta_R) E + (\beta_L^2 \hspace{-2pt}+\hspace{-2pt} \beta_R^2)\DeltaE^2 \hspace{-1pt}-\hspace{-1pt} \frac{\alpha_S N}{\mathcal{J}}\frac{2\pi^2(\beta_L\hspace{-2pt}+\hspace{-2pt}\beta_R)}{\beta_{\text{aux}}^2}\right]
\ee
where the factor of $T$ out front comes from an integral over $\Deltat$. In this expression, we have effectively reduced the $G,\Sigma$ variables to a single parameter $\beta_{\text{aux}}$ by putting all of the other directions in function space on shell. The saddle point of the final three-dimensional integral is at $\beta_L = \beta_R = 0$ with $\beta_{\text{aux}}$ fixed by
\be\label{sameasearlier}
E = \frac{\alpha_S N}{\mathcal{J}}\frac{2\pi^2}{\beta_{\text{aux}}^2}.
\ee
The expression on the RHS is the energy of the Schwarzian degree of freedom (see e.g. \cite{Maldacena:2016upp}) as a function of $\beta_{\text{aux}}$, and the saddle point condition fixes $\beta_{\text{aux}}$ by setting this equal to the argument $E$ of $Y_{E,\DeltaE}(T)$.

We can also see how adding an extra $L$-$R$ coupling can stabilize $\beta_{\text{aux}}$ without going to the microcanonical ensemble. A small value of $\lambda$ will be sufficient, so we can compute its effect by first order perturbation theory, simply evaluating the term
\begin{align}
I&\supset-\lambda \frac{J^2}{q}\big(1 + \frac{\beta^2}{T^2}\big)\int_0^T\int_0^T dt dt'\left(iG_{LR}(t,t')\right)^q \approx -\lambda \frac{J^2}{q}\big(1 + \frac{\beta^2}{T^2}\big)T\int_{-\infty}^\infty dt\left(iG_{LR}^{(\beta_{\text{aux}})}(t)\right)^q\notag\\
&\approx -\frac{q-2}{q^2}\tan(\frac{\pi}{q}) \frac{\lambda T}{\beta_{\text{aux}}}.\label{lambda}
\end{align}
In going to the second line we dropped corrections proportional to $\beta^2/T^2$. Adding (\ref{sch}) and (\ref{lambda}) together, we find a stable minimum at $\beta_{\text{aux}} \propto \frac{\beta}{J T\lambda}$.

\subsection{Interpolating between Brownian and regular SYK}
One can interpolate between Brownian and regular SYK by giving the couplings $J$ a correlation in time of the form:
\be
\langle J_{a_1\dots a_q}(t)J_{a_1'\dots a_q'}(t')\rangle = \delta_{a_1a_1'
}\dots \delta_{a_qa_q'}\,\frac{(q{-}1)!}{N^{q-1}}J^2(t-t'),
\ee
where the function $J^2(t)$ is intermediate between a delta function and a constant function. For the problem $\langle |Z(iT)|^2\rangle$, it is natural to define the function $J^2(t)$ to respect the periodicity of the time circle. For example, one can define
\be
J^2(t) = \left(\frac{J}{\sqrt{2\pi}\,t_J} + \frac{J^2}{2}\right)\left[\exp\left(-\frac{t^2}{2t_J^2}\right) + \exp\left(-\frac{(t-T)^2}{2t_J^2}\right)\right], \hspace{20pt} 0<t<T,
\ee
and periodically extend this function outside the range. The parameter $t_J$ determines the timescale over which the couplings are correlated.  We have Brownian SYK for $t_J\rightarrow 0$ and regular SYK for $t_J \rightarrow \infty$.

To study $\langle |Z(iT)|^2\rangle$ for this theory, we can use the $G,\Sigma$ action in (\ref{GSigmaSYK}), but with the substitution $J^2 \rightarrow J^2(t-t')$. One also replaces $J^2\rightarrow J^2(t)$ in the saddle point equations (\ref{saddlePt}). These equations must be solved numerically for generic $t_J$, but in the Brownian limit $t_J \rightarrow 0$ they are easy to solve by hand. The simplification is because $\Sigma_{ij}$ becomes short-ranged in time, due to the factor of $J^2$ in (\ref{saddlePt}) that gets replaced by $J^2(t)$. Indeed, $\Sigma_{LL}, \Sigma_{RR}$ become equal to zero because $J^2(t)$ multiplies an odd function of time. $\Sigma_{LR}(t)$ becomes proportional to a delta function. The solutions are (for $|t-t'|<T$)
\begin{align}\label{twoptbrowna}
G_{LL}(t,t') &= G_{RR}(t,t') = \frac{\text{sgn}(t-t')}{2}e^{-\mu|t-t'|} + \frac{\sinh(\mu(t-t'))}{1 + e^{\mu T}}\\
G_{LR}(t,t') &= -G_{LR}(t,t')  = \pm i \left[\frac{1}{2}e^{-\mu|t-t'|} - \frac{\cosh(\mu(t-t'))}{1 + e^{\mu T}}\right]\label{twoptbrownb}
\end{align}
where the parameter $\mu$ is determined by requiring that it solves the equation
\be\label{mueq}
\mu = J\left(\frac{1}{2} - \frac{1}{1+e^{\mu T}}\right)^{q-1}.
\ee
Here we have written the solutions with both time arguments explicit in order to make the following point. In section \ref{brownsyk}, we were able to analyze the Brownian model using a smaller set of variables, only $G_{LR}(t,t)$ and $\Sigma_{LR}(t,t)$ with both times equal. One can derive this description starting from (\ref{GSigmaSYK}) by noting that when $J^2(t)$ is proportional to a delta function, the $G_{ij}(t,t')$ variables (other than $G_{LR}(t,t)$) enter the action only through the $G_{ij}(t,t')\Sigma_{ij}(t,t')$ term. Integrating over these $G_{ij}(t,t')$ variables sets the corresponding $\Sigma_{ij}(t,t')$ variables to zero and we end up with (\ref{GSigma}), after redefining $\Sigma_{LR}$ by a factor of two. The present description with $G_{ij}(t,t')$ is not as economical, but it gives a little more information, since (\ref{twoptbrowna}) and (\ref{twoptbrownb}) give the full time dependence of the correlators.

As a second point, note that these solutions are isolated; we don't have $\beta_{\text{aux}}$ and $\Deltat$ zero modes. We expect that this remains true for any finite value of $t_J$, and that the zero modes only appear in the regular SYK limit of $t_J\rightarrow \infty$. A simple guess is that the quadratic action for $\Deltat$ is proportional to $N J T \Deltat^2/t_J^2$. For very large $t_J$ we can ignore this action, and the integral over $\Deltat$ will give a $T$-linear ramp. But as $T$ grows, the action will become significant, eventually cutting off the $\Deltat$ integral, and presumably leading to a non-monotonic $\langle |Z(iT)|^2\rangle$ that decays, approaching an order one value at late time.

Finally, one reason to discuss this interpolation is the following. In the Brownian SYK we can analyze the $G,\Sigma$ path integral exactly, see appendix \ref{exactapp}, and we know that the nontrivial saddle point contributes to the path integral. Via the interpolation, one can connect the ramp saddle points in regular SYK to this one from Brownian SYK. This gives us some confidence (beyond the fact that the answer is reasonable) that these ramp saddle points should indeed contribute to the path integral for regular SYK.

\section{Gravity}\label{sec:gravity}
We would now like to understand analogous saddle points in a simple gravity context. This section can be read independently of section \ref{SYK}, but our motivation for the specific gravity solutions below came from trying to match the SYK configurations in section \ref{conformalDetails}. 

\subsection{JT gravity}
The theory we consider first is the Jackiw-Teitelboim theory of gravity in two dimensional AdS space. The action is
\be\label{JTaction}
I_{JT} = -\frac{\phi_0}{2}\left[\int \sqrt{g}R + 2\int_{bdy}\sqrt{h}K\right] - \frac{1}{2}\left[\int \sqrt{g}\,\phi(R+2) + 2\phi_b\int_{bdy}\sqrt{h}K\right].
\ee
The equation of motion we get by varying with respect to $\phi$ imposes that $R = -2$, which in two dimensions is enough to locally fix the geometry to be a piece of AdS${}_2$. The problem is to figure out which piece we are supposed to choose in order to compute the quantity $Z(\beta+iT)Z(\beta-iT)$. One choice would be to take two disconnected geometries corresponding to ``Euclidean'' black holes with inverse temperature $\beta+iT$ and $\beta-iT$. Each solution has the topology of the disk, and in total we have two disconnected disks. Such a solution is analogous to the trivial $G_{LR} = 0$ case in the SYK discussions above. This contribution is the gravity description of the ``slope'' region, and it decays to zero for large $T$. 

A more interesting choice is a connected geometry, with the topology of a cylinder that we will refer to as the ``double cone.'' We will first describe this solution very naively. The simplest Lorentzian solution in JT gravity corresponds to a two-sided AdS${}_2$ black hole, which is the same thing as AdS${}_2$ in Rindler coordinates. The metric and dilaton are
\be\label{phih}
ds^2 = -\sinh^2(\rho) d\t^2 + d\rho^2, \hspace{20pt} \phi = \phi_h \cosh(\rho).
\ee
This reference solution has a time translation symmetry under shifts in $\t$. Roughly, the solution we are interested in is an identification of this geometry by $\t \sim \t + \T$. This corresponds to identifying the Rindler region by a boost. Since the boost is a forwards Lorentzian time translation at the $R$ boundary and a backwards one at the $L$ boundary, this identification will impose the correct periodicity to contribute to $Z(iT)Z(-iT)$. The geometry can be visualized as a type of ``double cone'' with closed timelike curves, and with the two tips of the cones meeting at $\rho = 0$ where we have a fixed point of the identification. There are naturally two parameters associated with this type of solution. One is the relationship between $\T$ and $T$, which we will see is related to the $\beta_{\text{aux}}$ parameter discussed above, or equivalently $\phi_h$ in (\ref{phih}). The other is a choice of where to place the origin of time coordinates on the $L$ and $R$ boundaries. Because of the time-translation symmetry of the identified geometry, simultaneous translations of the origin on the $L$ and $R$ have no effect, but a relative translation is meaningful. In other words, as with the SYK solutions, the double cone spontaneously breaks the independent $L$ and $R$ translation symmetries, leaving a compact zero mode $\Deltat$ with volume proportional to the period $T$. Such a configuration is therefore a candidate for explaining the ramp.

\begin{figure}[ht]
\begin{center}
\includegraphics[width = .7\textwidth]{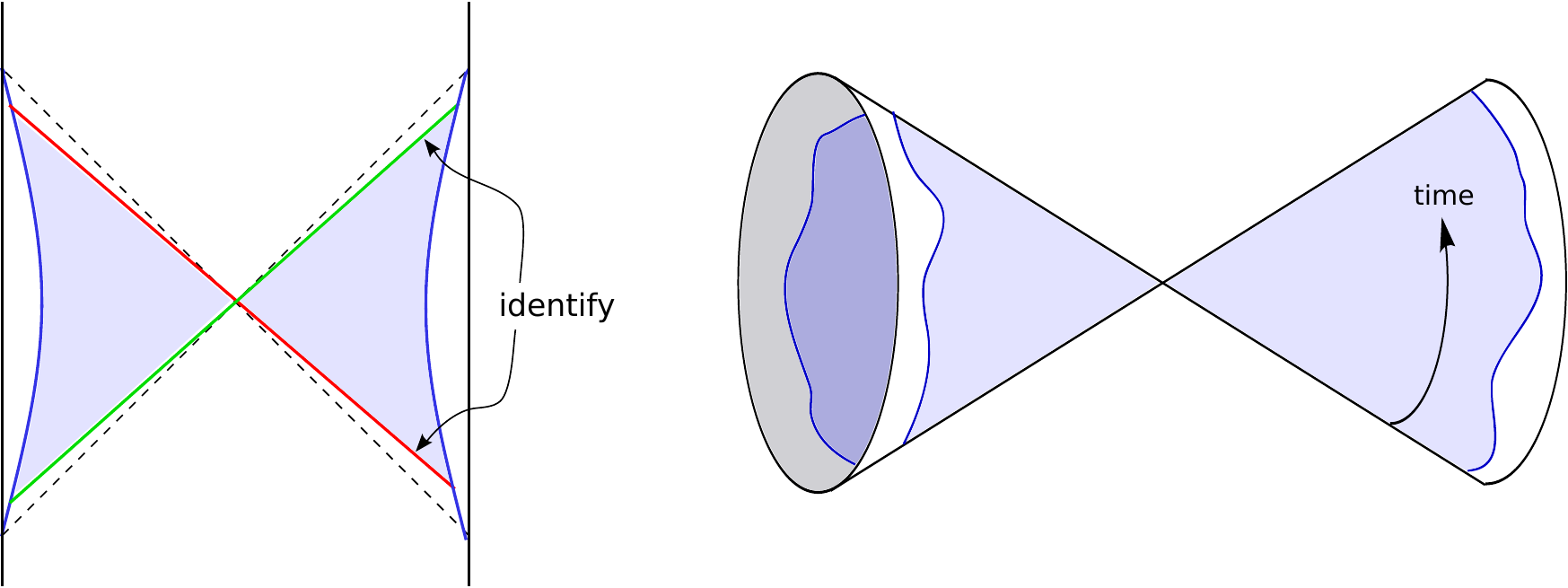}
\caption{{\small The double cone. At left we indicate two identification surfaces in AdS${}_2$ (red and green, indicated by arrows). The blue curved lines represent the regulated boundaries. At right we have folded the geometry into a double cone. We also made the regulated boundary wiggly to represent the boundary graviton degree of freedom.}}\label{figdoubleCone}
\end{center}
\end{figure}

\begin{figure}[t]
\begin{center}
\includegraphics[width = .5\textwidth]{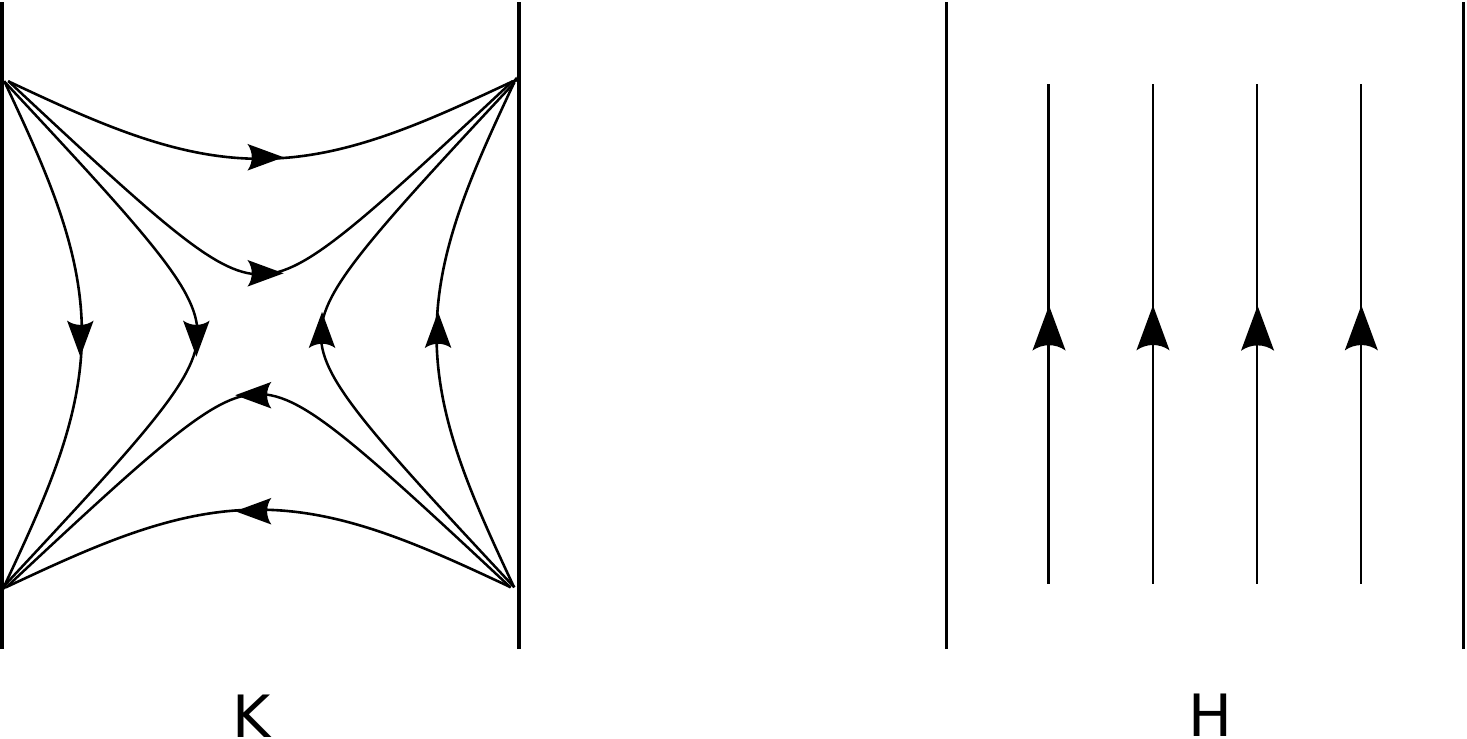}
\caption{{\small The vector fields associated to the $SO(1,2)$ generators $K$ and $H$. The diagrams are global AdS${}_2$, with time running vertically and the two boundaries shown.}}\label{figHandK}
\end{center}
\end{figure}

We will now describe the double cone solutions more systematically. We will start by considering $Z(\beta+iT)Z(\beta-iT)$, and we will see how to fix the $\beta_{\text{aux}}$ instability below. The $L$ and $R$ boundaries of the cylinder are at separate asymptotic boundaries in AdS${}_2$. The circle that forms the boundary on the $L$ side of the cylinder should correspond to periodic identification by Euclidean time $\beta+iT$, and on the $R$ side it should correspond to $\beta-iT$. Motivated by the form of the solutions in SYK, we would like to consider the case where the geometry has a $U(1)$ time translation symmetry. We can guarantee such a symmetry while also satisfying the periodicity requirements at the two boundaries by choosing a new time coordinate $\t$ for AdS${}_2$ that corresponds to the generator
\be\label{combine}
\partial_{\t} = iK -\frac{\beta}{T}H = -\partial_{t_{Rindler}} - i\frac{\beta}{T}\partial_{t_{global}}
\ee
and then periodically identifying the geometry by $\t \sim \t+\T$, where as we will see $\T$ should be proportional to $T$. Here $K$ is the generator of Rindler time translations, and $H$ is the generator of global time translations, see figure \ref{figHandK}. The point of taking this type of combination is that $K$ translates time in opposite ways on the two boundaries, and $H$ translates them in the same direction, so periodicity with respect to $\t$ implies periodicity proportional to $\beta\pm iT$ on the two boundaries.

To describe the space in these coordinates more explicitly, it is convenient to use embedding coordinates for AdS${}_2$, which we will write in the Rindler and global coordinate systems:
\begin{align}
Y_{-1} &= \cosh(\rho) \hspace{64pt}= \cosh(r) \cos(t_{global})\notag\\
Y_0 &= \sinh(\rho)\sinh(t_{Rindler}) = \cosh(r)\sin(t_{global})\label{coords}\\
Y_1 &= \sinh(\rho)\cosh(t_{Rindler}) = \sinh(r)\notag
\end{align}
A convenient basis for the group of isometries $SO(1,2)$ is
\be\label{matrices}
H = \left(\begin{array}{ccc} 0 & -i & 0 \\ i & 0 & 0 \\ 0 & 0 & 0\end{array}\right),\hspace{20pt}
K = \left(\begin{array}{ccc} 0 & 0 & 0 \\ 0 & 0 & i \\ 0 & i & 0\end{array}\right),\hspace{20pt}
P = \left(\begin{array}{ccc} 0 & 0 & i \\ 0 & 0 & 0 \\ i & 0 & 0\end{array}\right).
\ee
We can write points in the new coordinate system with $\t$ as a time coordinate by writing
\be\label{exponential}
\left(\begin{array}{c}Y_{-1} \\ Y_0 \\ Y_1\end{array}\right)=\exp\left(\t\partial_{\t}\right)\left(\begin{array}{c}\cosh(\rho) \\ 0 \\ \sinh(\rho)\end{array}\right) = \exp\left[\widetilde{t}\cdot\left(\begin{array}{ccc} 0 & i\frac{\beta}{T} & 0 \\ -i\frac{\beta}{T} & 0 & -1 \\ 0 & -1 & 0\end{array}\right)\right]\left(\begin{array}{c}\cosh(\rho) \\ 0 \\ \sinh(\rho)\end{array}\right).
\ee
To compute the metric in these coordinates we use the metric induced from the embedding space $ds^2 = -dY_{-1}^2 - dY_0^2 + dY_1^2$. Since $\t$ corresponds to a symmetry direction we can compute this for small $\t$, linearizing the exponential in (\ref{exponential}). One finds
\be\label{metric}
ds^2 = -\left(\sinh(\rho) +\frac{i\beta}{T}\cosh(\rho)\right)^2d\t^2 + d\rho^2, \hspace{20pt} \t \sim \t+\T.
\ee
Note that in this metric, the singularity at $\rho = 0$ in the naive version discussed above is resolved by the $i\beta/T$ term in the metric. We will come back to this point below.

There is one further parameter we need to specify in the metric (\ref{metric}), which we can think about as the relationship between the time $\t$ and the time $t$. Let's briefly review what these variables mean. We are studying $Z(\beta+iT)Z(\beta-iT)$. This means periodicity in Euclidean time $\tau \sim \tau + \beta+iT$ on the $L$ system, and $\tau \sim \tau + \beta-iT$ on the $R$ system. We use the notation $t$ to define a complex time coordinate that in both cases runs from zero to $T$.  So, on the $L$ system we have $\tau = (\frac{\beta}{T}+i)t$ and on the $R$ system we have $\tau = (\frac{\beta}{T}-i)t$. Now, the idea is that near the boundary, the bulk time coordinate $\t$ is proportional to our boundary time coordinate $t$. This coefficient of proportionality depends on exactly where we put the cutoff surface, and the freedom here will correspond to the parameter $\beta_{\text{aux}}$ in our discussion of SYK.

To make this precise, we introduce a holographic renormalization parameter $\epsilon$ and we relate the boundary proper Euclidean time $\tau$ to the bulk metric via
\be
d\tau^2_{boundary} = \epsilon^2 ds^2_{bulk}\big|_{\rho = \pm \rho_c}.
\ee
This gives a relationship between $\tau$ and $\t$ on the two boundaries. When we translate both of these to relationships between $t$ and $\t$, we find in both cases
\be\label{cutoff}
t = \frac{\epsilon e^{\rho_c}}{2}\widetilde{t} = \frac{\beta_{\text{aux}}}{2\pi}\t.
\ee
Anticipating slightly, in the second equality we have given an interpretation to the factor $\epsilon e^{\rho_c}/2$ as being equal to $\beta_{\text{aux}}/(2\pi)$. To justify this relationship we can compute geodesic distances between boundary points, and for this numerical factor we find agreement between the geodesic approximation and the expressions in (\ref{GLL}) and (\ref{GLR}).

It is interesting to compute the JT action (\ref{JTaction}) for the space (\ref{metric}). The term proportional to $\phi_0$ is a topological term, which is normally responsible for contributing the zero-temperature entropy $S_0$. However, in our case it gives zero because our space has the topology of a cylinder, which has zero Euler characteristic. This is in keeping with expectations (e.g. from SYK) that the value of the ramp should {\it not} be proportional to $e^{S_0}$. The bulk term in the action proportional to $\phi(R+2)$ gives zero because $R + 2 = 0$ for the metric (\ref{metric}), so all that remains is the boundary term $\phi_b\int \sqrt{h} K$. The boundary condition for $\phi$ is that $\phi_b = \frac{\phi_r}{\epsilon}$, where $\phi_r$ is a renormalized boundary value of the dilaton. The extrinsic curvature scalar is of order one, but the length of the curve and the value $\phi_b$ are large for large cutoff radius, so the expression is divergent. Renormalizing by subtracting a multiple of $\phi_b$ times the length of the boundary curve, and adding the contributions from the two boundaries, we find the finite leftover
\be\label{JTactionans}
I_{JT} =\phi_r\frac{2\beta}{T}\frac{\T}{\epsilon e^{\rho_c}} =\phi_r\frac{4\pi^2\beta}{\beta_{\text{aux}}^2}.
\ee
This is the same as what we got from the Schwarzian action (\ref{sch}). Of course, this follows from the equivalence of JT to the Schwarzian \cite{Maldacena:2016upp,Jensen:2016pah,Engelsoy:2016xyb}.

Note that the action (\ref{JTactionans}) is not stationary with respect to $\beta_{\text{aux}}$, which is a parameter of the configuration. What this means is that we do not have a true solution to the equations of the JT theory. This can be understood as follows. In addition to the equation $R +2=0$ that we get by varying $\phi$, there is a second equation that we get by varying the metric. This can be written as $T^\phi_{\mu\nu} = 0$, where
\be
T^{\phi}_{\mu\nu} = \nabla_\mu\nabla_\nu\phi + (\phi - \nabla^2\phi)g_{\mu\nu}.
\ee
To have a solution with time translation symmetry, we would like to impose that $\phi$ is a function of $\rho$ only, $\phi(\rho)$. With this assumption, the nonzero components are
\be
T^{\phi}_{\widetilde{t}\widetilde{t}} = \left(\sinh(\rho) + \frac{i\beta}{T}\cosh(\rho)\right)^2(\phi'' - \phi) ,\hspace{20pt} T^\phi_{\rho\rho} =  \phi -\frac{\cosh(\rho) + \frac{i\beta}{T}\sinh(\rho)}{\sinh(\rho) + \frac{i\beta}{T}\cosh(\rho)}\phi'.
\ee
We are supposed to solve these equations together with the boundary condition that $\phi$ takes the same boundary value $\phi_b$ at the $L$ and $R$ boundaries $\rho = \pm \rho_c$. Then $\phi$ should be an even function of $\rho$ and setting $T^\phi_{\t\t} = 0$ implies that $\phi \propto \cosh(\rho)$. However, this is inconsistent because now $T^\phi_{\rho\rho}$ is nonzero, and proportional to $\beta/T$. So in fact there is no solution of the full problem. 

This is the dilaton gravity version of the phenomenon that we saw in the SYK case above, that for nonzero $\beta$ we don't have a solution of the full equations because of a runaway pressure on $\beta_{\text{aux}}$. As there, we can deal with this by going to the microcanonical quantity $Y_{E,\DeltaE}(T)$ (\ref{Y}). This amounts to allowing $\beta = \frac{\beta_L+\beta_R}{2}$ to be determined dynamically, but also imposing stationarity with respect to its variation. Requiring that we have an on-shell solution for the dilaton imposes that $\beta = 0$. To impose stationarity with respect to $\beta$, we have to cancel the $\beta$ dependence of the explicit factor $2\beta E$ against $-I_{JT}$ with (\ref{JTactionans}). This gives
\be
2E = \phi_r\frac{4\pi^2}{\beta_{\text{aux}}^2}
\ee
which fixes $\beta_{\text{aux}}$. This is the same condition as (\ref{sameasearlier}) after translating to SYK conventions.\footnote{In section \ref{nonzerobeta} we also discussed the possibility of stabilizing $\beta_{\text{aux}}$ by adding a term to the action
\be
I\supset -\lambda \int_0^T\int_0^T dt dt'\mathcal{O}_L(t) \mathcal{O}_R(t')
\ee
where $\mathcal{O}_L,\mathcal{O}_R$ are operators on the $L$ and $R$ boundaries. The bulk effect of this should be related to the eternal wormhole effect described by Maldacena and Qi \cite{Maldacena:2018lmt}, where a two-sided coupling leads to stress energy in the bulk that allows the equation for the dilaton to be solved even with nonzero $\beta$.}

\subsection{More general black holes}
In writing down the double cone configuration, we were pretty much just translating the SYK saddle points from section \ref{conformalDetails} into bulk Jackiw-Teitelboim variables.

However, we can now generalize this strategy to higher dimensional stationary black holes, finding a contribution to $|Y_{E,\DeltaE}(T)|^2$ defined in (\ref{Y}). Let's briefly describe the gravity boundary conditions for computing this quantity. It's helpful to start by considering the quantity $Z(\beta_L)Z(\beta_R)$. Here we study a geometry with two asymptotic boundaries, and with periodicity in the Euclidean time direction proportional (up to a holographic renormalization parameter) to $\beta_L$ and $\beta_R$ on the two boundaries. This procedure is most familiar for real $\beta_L$ and $\beta_R$, but it makes sense for complex values as well. In order to compute $|Y_{E,\DeltaE}(T)|^2$ by saddle point, we allow $\beta_{L,R}$ to vary, and look for solutions to the bulk equations of motion together with the saddle point conditions for $\beta_{L,R}$:
\begin{align}\label{cond1}
E + 2\beta_L \DeltaE  + \partial_{\beta_L}\log\left[Z(\beta_L+iT)Z(\beta_R-iT)\right] &= 0\\
E + 2\beta_R \DeltaE  + \partial_{\beta_R}\log\left[Z(\beta_L+iT)Z(\beta_R-iT)\right] &= 0.\label{cond2}
\end{align}
We will discuss a simple family of solutions to these equations, where $\beta_L = \beta_R = 0$ and the $\DeltaE$ terms drop out. The solution should be periodic by Lorentzian time $\pm T$ in the two asymptotic regions. Since $\beta_L$ is conjugate to the asymptotic energy in the $L$ region, and similarly for $R$, the two conditions (\ref{cond1}), (\ref{cond2}) imply that the energy of the $L$ and $R$ asymptotic regions should both be equal to $E$.

\begin{figure}[t]
\begin{center}
\includegraphics[width = .4\textwidth]{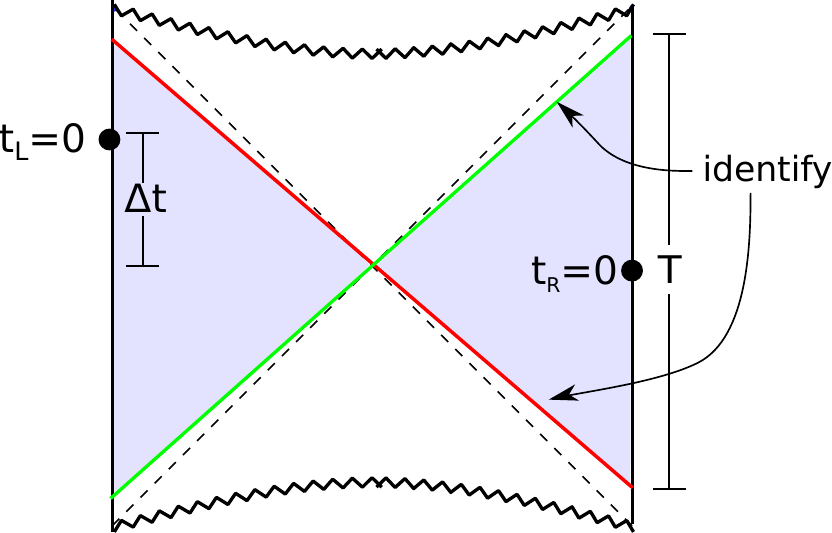}
\caption{{\small The identification we consider for a higher dimensional black hole cuts out the shaded blue region, which forms a double cone. The green and red identification surfaces differ by Schwarzschild time $T$. The zero mode that leads to the ramp factor of $T$ is the freedom to insert a relative shift $\Deltat$ between the Schwarzschild time coordinates of the the origin of time for the $L$ and $R$ theories.}}\label{figdoubleCone2}
\end{center}
\end{figure}

To construct a solution, one can start with the thermofield double black hole with energy $E$ in real time. One then periodically identifies the metric by the Killing (e.g. Schwarzschild) time $t\sim t+T$. Since this time variable runs forwards on the $R$ boundary and backwards on $L$, the periodicities on the two boundaries will be by $\pm T$ in Lorentzian time. The resulting geometry is a type of Lorentzian double cone, see figure \ref{figdoubleCone2}.
Of course, this identification has a fixed point at the horizon, which makes a singularity. At least in bulk effective field theory, it can be treated by going to Rindler coordinates near the horizon and deforming $\rho$ into the upper half plane slightly, as we discussed above in the JT gravity case.

The classical action of this solution is zero. One way to see this is that because the tip of the double cone is smooth in the sense described in the next subsection, there is no contribution from there. The Einstein-Hilbert action and Gibbons-Hawking boundary terms on the $L$ and $R$ sides are both pure imaginary since we have a real geometry in Lorentzian signature, and they cancel between the two sides. 

Importantly, there will be a factor of $T$ associated to this solution, from the integral over a compact zero mode $\Deltat$. The interpretation of this zero mode in gravity is as follows. We keep exactly the same geometry, but we change where the origin of coordinates for the $L$ and $R$ theories sit. If we advance both in the same direction (in Killing time), then because of the $U(1)$ symmetry of the solution we will not have done anything at all. But if we move them in opposite directions, then we have a new solution. The parameter $\Deltat$ is defined as the difference in Killing times of the points $t_L = 0$ and $t_R = 0$, see figure \ref{figdoubleCone2}.

The solution described here, and the fact that it has zero action, is quite general. In particular, such a periodic identification can be made for any stationary black hole, including charged and/or rotating ones. So in computing $|Y_{E,\DeltaE}(T)|^2$, we would have to sum over black holes with all values of angular momentum and gauge charges, subject to the constraints from extremality given the energy $E$. Since the action is zero, no particular black hole will dominate the sum. Instead, we expect the sum to contribute a numerical factor that multiplies the ramp function. This is analogous to the two solutions that we found in SYK, with $G_{LR}$ either positive imaginary or negative imaginary. There the interpretation was a separate contribution to the ramp from the two $(-1)^F$ symmetry sectors. Here there are separate contributions from all global symmetry charge sectors, including rotations and internal global symmetries. (In order to focus on just one of these solutions, one could of course define a variant of $Y_{E,\DeltaE}(T)$ at fixed global symmetry charge.)

\subsection{Fluctuating fields on the double cone}
In addition to the classical action, we should consider the effect of fields fluctuating about this saddle point. The boundary fluctuations of pure JT gravity are mild and are analyzed in appendix \ref{JTexact}. However, in a bulk theory with propagating fields, it seems to be subtle. The saddle points for $Y_{E,\DeltaE}(T)$ have $\beta = 0$, which means a space with purely Lorentzian periodicity. In this setting quantum field theory is not obviously well defined. In addition, the space looks singular due to the fixed point of identification at $\rho = 0$. It is very possible that a general bulk theory cannot be defined on this space (actually, this would be a good thing, see Discussion). However, at least in some examples, the partition function of quantum fields seems surprisingly well behaved; the double cone manages to avoid obvious perturbative problems one would expect given the Lorentzian periodicity and $\rho = 0$ singularity.

As one example, we can consider AdS${}_2$. Here the $SO(1,2)$ or equivalently $SL(2,\mathbb{R})$ symmetry of quantum fields on a fixed AdS${}_2$ space makes the argument easy. In defining the double cone, we are identifying AdS${}_2$ by the action of the complexified $SO(1,2)$ element $e^{iK\T - \widetilde{\beta} H}$, where $\widetilde{\beta} = \beta \T/T$. The partition function of quantum fields on the periodically identified space can therefore be computed in the Hilbert space formalism as $\Tr(e^{iK\T - \tilde{\beta} H})$ where the trace is over the Hilbert space of the bulk fields on a global slice through AdS${}_2$. Now, a useful fact that one can check from the explicit matrix representation (\ref{matrices}) is that
\be
e^{-aP}(iK \T - \widetilde{\beta} H)e^{aP} = -\sqrt{\T^2 + \widetilde{\beta}^2}\,H, \hspace{20pt} \tan(a) = \frac{T}{\beta}.
\ee
So the generator we are identifying by is conjugate, by a complexified $SO(1,2)$ group element, to a multiple of $H$. If the bulk fields have an exact $SO(1,2)$ symmetry, so that the trace is over a sum of $SO(1,2)$ representations, then it follows that $\Tr(e^{iK\T - \widetilde{\beta} H}) = \Tr(e^{-\sqrt{\T^2+\widetilde{\beta}^2}H})$. As $\widetilde{\beta} \rightarrow 0$ we find the partition function of fields in global AdS${}_2$ at inverse temperature $\T$. This is well behaved: in fact for large times $\T$, the quantum fields are effectively projected into the ground state, and the partition function approaches one. Note that in this way we can give a meaning to $\Tr(e^{iKT})$, whereas $\Tr(e^{iHT})$ would not be well defined. For example, for a free scalar field dual to an operator of dimension $h$, we have
\be\label{naturalboundary}
\Tr(q^H) = \prod_{n = 0}^\infty \frac{1}{1-q^{h+n}}.
\ee
To study $\lim_{\epsilon\rightarrow 0}\Tr(e^{iHT-\epsilon H})$ we would take $q \rightarrow e^{iT}$, but (\ref{naturalboundary}) has a ``natural boundary'' at the unit circle that makes this undefined. On the other hand, for $\lim_{\epsilon\rightarrow 0}\Tr(e^{iKT-\epsilon H})$, we end up with the same function but with $q = e^{-T}$ well inside the unit disk.

One can also think about perturbation theory on a double cone spacetime. A potential problem is the region where vertices are integrated near the tip $\rho = 0$. Since this is a fixed point of the identification, the sum over images to define the propagator diverges there and the propagator is singular. However, it seems that one can deform the contour for $\rho$ to avoid this point. In a Rindler approximation near the horizon, we write the metric as
\be\label{witheps}
ds^2 = -(\rho + i\epsilon)^2dt^2+d\rho^2= (\epsilon-i\rho)^2dt^2+d\rho^2, \hspace{20pt} t\sim t+T.
\ee
The $i\epsilon$ prescription can be motivated from the contour of integration for $\beta = \frac{\beta_L+\beta_R}{2}$ in the microcanonical transform (\ref{Y}), which requires that we approach the saddle point at $\beta = 0$ from a direction where $\beta$ has a positive real part. Then we can view the $\beta$ parameter in e.g. (\ref{metric}) as a positive infinitesimal. Now, the idea is that the contour for $\rho$ can be deformed into the upper half plane, avoiding the singularity at $\rho = 0$. So perturbation theory seems to be well behaved. Note that this deformation would not be possible for a different type of regularized double cone:
\be\label{traditionalCone}
ds^2 = -(\rho^2 + \epsilon^2)dt^2 + d\rho^2, \hspace{20pt} t\sim t+T.
\ee
In this case, the defining contour for $\rho$ along the real axis is caught between two zeros of $\rho^2 + \epsilon^2$, and it cannot be usefully deformed. Indeed, as $\epsilon$ vanishes (\ref{traditionalCone}) is unambiguously singular: the curvature diverges at $\rho = 0$. For (\ref{witheps}) the curvature is exactly zero everywhere.

The two spaces (\ref{witheps}) and (\ref{traditionalCone}) also illustrate how the double cone can avoid the problem associated to Lorentzian periodicity. Consider a massless field with Dirichlet boundary conditions $\phi=0$ at $\rho = \pm 1$. One can evaluate the partition function by decomposing in modes in space and treating each mode as a harmonic oscillator with some frequency. Away from the tip, (say for $\rho>0$) candidate mode solutions with frequency $\omega$ would be
\be
\phi_\omega(\rho) = A\rho^{i\omega} +B \rho^{-i\omega}.
\ee
The boundary condition at $\rho = 1$ requires $B = -A$. We now consider the boundary condition at $\rho  = -1$. For the space (\ref{witheps}), when we continue to negative $\rho$, we find
\be
\phi_\omega(\rho) = A e^{-\pi \omega}(-\rho)^{i\omega} -Ae^{\pi \omega}(-\rho)^{-i\omega}.
\ee
To have $\phi$ vanish at $\rho = -1$ we need $e^{\pi\omega} = \pm 1$ which gives imaginary frequencies and suppression at large $T$. On the other hand, for the space (\ref{traditionalCone}), the boundary conditions can be satisfied with real frequencies, leading to wild oscillations in $T$ from large $\omega$ modes.

To be clear: we are not claiming that the fluctuations are under control in general. In fact, we hope that they are not (see Discussion). It's just that we haven't found a pathology.

{\bf Added in v2:} in the above, we argued that matter fields do not make a large contribution for large $\widetilde{T}$. But in principle the integral over $\widetilde{\beta}$ includes a region where $\widetilde{T} = T\widetilde{\beta}/\beta$ is small. The integral over this region would actually be divergent in the probe approximation with a free field. It is presumably regulated in a more complete bulk theory such as the dual to SYK. See section 6.1 of \cite{Saad:2019lba} for further comments.

\section{Discussion}
\subsection{Correlation functions}

 So far we have mostly discussed the spectral form factor and the related quantity $|Y_{E,\DeltaE}(T)|^2$. However, we expect certain correlation functions, such as (\ref{thermalcorr}) to share the same general ramp-plateau structure \cite{Cotler:2016fpe}. We expect that the basic origin of the ramp is the same. Instead of the ``wormhole'' connecting two disconnected systems, it connects the two sides of a timefold, with the linear growth of the ramp coming from a shift in how these two halves of the timefold are connected. Note that in this case the time translation symmetry leading to the factor of $T$ would have to be approximate, not exact.

 Consider the two-sided correlator from (\ref{thermalcorr}):
    \be
    f_\beta(T) = \frac{1}{Z(\beta)}\sum_{n,m} |\langle n | O |m\rangle|^2e^{-(\frac{\beta}{2}+i T)E_n} e^{-(\frac{\beta}{2}-i T)E_m} 
  \ee
For reasonably large times the sum will be dominated by energies that are close together. Within a given energy band, ETH predicts that the matrix element squared will have a typical value of order $e^{-S(E)}$. Including this factor and comparing to (\ref{rmtramp}), the expected ramp and plateau behavior is
\be
f_{\beta}(T) \sim \frac{1}{Z(\beta)}\int dE\, e^{-\beta E - S(E)}\text{min}(T,e^{S(E)}).
\ee
Omitting the normalization factor $Z^{-1}$, the integrand is proportional to $e^{-\beta E - S(E)}$, which includes a factor of $e^{-S_0}$. In JT gravity, we could get such a contribution from a space that is topologically a handle attached to a disk, which has Euler characteristic $\chi = -1$. It would be interesting to find candidate saddle points in SYK and in gravity.

\subsection{Wiggles and factorization}\label{singlesample}
An important property of the spectral form factor is that the ramp and plateau are not self-averaging \cite{prange1997spectral}. This is evident in figure \ref{fig:g-SYK}. So, for a fixed Hamiltonian system, even though we are summing over many energy levels, the result should be a function with $O(1)$ fluctuations. The smooth ramp can be made visible by a time average or a disorder average, but the exact function should be erratic.

In this paper we have presented saddle points that give a smooth ramp, not an erratic one. This is perfectly reasonable for the SYK model, since in the $G,\Sigma$ formulation we are doing a disorder average, which washes out the wiggles and leaves a smooth function. So we have some confidence that the solutions described here correctly describe the ramp in SYK. However, we also discussed ``double cone'' solutions for more general gravity theories, and found only a smooth ramp contribution from these. This could be the correct answer in JT gravity, for which the partition function does not have an interpretation as trace in a Hilbert space \cite{Stanford:2017thb,Harlow:2018tqv} (see \cite{Maloney:2007ud} for earlier related work in 3d), and which perhaps needs to be interpreted as a disorder-averaged theory.

It is interesting to consider the possibility that there could be more theories of quantum gravity like this. However, it is widely believed that in many cases, string theories are exactly dual to specific quantum systems (such as $\mathcal{N} = 4$ super Yang-Mills), with no disorder average. For such systems, the bulk theory should produce a wiggly ramp, not a smooth one. So, to put it plainly, we are getting the wrong answer. What is being left out?

One possibility is that in a bulk theory dual to a specific quantum system, the path integral for fluctuations about the double cone would be badly behaved due to contributions from high energy bulk states. Such states could give contributions that oscillate rapidly in time $T$. In this situation, the simple double cone saddle point might still give the correct answer for the time-average. In other words, some pathology of the double cone could produce the wiggles, but time averaging would tame it. This needs more study.

The simple gravity saddle points do seem to get one aspect of the wiggles right. For $|Y_{E,\DeltaE}(T)|^{2k}$, one can construct $k!$ different solutions, where the $k$ copies of $Y$ are paired by separate double cones to some permutation of the $k$ copies of $Y^*$. This coincides with the expected behavior of ensemble averages of $|Y_{E,\DeltaE}(T)|^{2k}$ in GUE random matrix theory, where
\be
\langle |Y_{E,\DeltaE}(T)|^{2k}\rangle \approx k!\langle |Y_{E,\DeltaE}(T)|^2\rangle^k.
\ee

The noise is crucial here: if $|Y_{\DeltaE,E}(T)|^2$ were self-averaging, then the expectation value would approximately factorize, and there wouldn't be a factor of $k!$. The ensemble average is also essential. If we remove the angle brackets representing either time average or disorder average, this expression doesn't make sense. It seems like gravity wants to give the answer for some kind of ensemble average, including the correct statistics for the noisy fluctuations within the ensemble.\footnote{We expect that other statistics, like time autocorrelation functions, are also given by the gravity saddles.}

This reinforces a more basic point, that the double cone does not respect factorization: $|Y_{E,\DeltaE}(T)|^2$ is a product of separate factors for the $L$ and $R$ systems, and the double cone represents a correlation between them.  Several of our colleagues have emphasized to us that this is reminiscent of the situation with Euclidean wormholes \cite{Coleman:1988cy,Maldacena:2004rf,ArkaniHamed:2007js}, for which the correct AdS/CFT interpretation remains unclear. One possibility is that Euclidean wormholes, and also the double cone, should not be included in the gravity path integral.\footnote{The failure of factorization in JT gravity was recently discussed in \cite{Harlow:2018tqv}, where it was argued that additional degrees of freedom need to be added to the bulk theory in order to restore factorization, see figure 1 of \cite{VanRaamsdonk:2010pw}. Perhaps these extra degrees of freedom (black hole microstates?) are needed in order to see the wiggles in a non-averaged system.}

\subsection{Toward the plateau}\label{toward}
To end this paper we return to the averaged quantity $\langle Z(iT) Z(-iT) \rangle$ and make some preliminary remarks about its asymptotic late time behavior -- the plateau.\footnote{Some related remarks about the SYK Brownian circuit are made in Appendix \ref{finiteL}.}   In random matrix theory the plateau has a very different origin than the ramp.   The spectral form factor is the Fourier transform of the eigenvalue pair correlation function which for GUE we can write schematically as
\begin{align}\label{firstline}
 \langle\rho(E) \rho(E') \rangle  &\sim  \frac{1}{L}\delta(E-E') +1 - \frac{\sin^2(L(E-E'))}{L^2(E-E')^2} \\
\langle Z(iT) Z(-iT) \rangle &= L^2 \int dE dE' e^{iT(E - E')}  \langle\rho(E) \rho(E') \rangle 
\end{align}

It is helpful to imagine using $\sin^2(x) = \frac{1}{2} - \frac{\cos(2x)}{2}$ to rewrite the numerator in (\ref{firstline}) as a constant piece plus something that averages to zero. The ramp arises from dropping the oscillatory $\cos(2L(E-E'))$ factor and Fourier transforming the leftover $\frac{-1}{L^2(E-E')^2}$ factor.  Note that this is of relative order $1/L^2$, perturbative in standard double line matrix perturbation theory for rank $L$ matrices.  Diagrams are weighted by $L^\chi$ where $\chi$ is their Euler character.  For $\langle Z(iT) Z(-iT)\rangle$ the leading diagrams have the topology of two disconnected discs ($\chi = 2$), which produce the early time slope contribution of order $L^2$.\footnote{The shape of the slope comes from the more accurate version of \eqref{firstline} with  $1 \rightarrow \langle\rho(E)\rangle \langle \rho(E')\rangle$ .} The ramp  is the first connected contribution which comes from ladder diagrams with the topology of a cylinder \cite{Brezin:1993qg}.  These have $\chi = 0$ so are of  order $L^0$.  The linear $T$ behavior comes from ladder diagrams where one side is cyclically permuted relative to the other.   The analogy to the double cone discussed in this paper is clear.
  
There are double line diagrams of all higher genus giving an (asymptotic) series in $1/L^2$, but because of cancellations  the long time behavior of the microcanonical version of $\langle Z(iT) Z(-iT) \rangle$, $\langle |Y_{E,\DeltaE}(T)|^2$,  only receives contributions from the cylinder, to all orders in $1/L^2$.  The plateau behavior comes from the oscillating factor $\cos(2L(E-E'))$ which is nonperturbative in $1/L^2$, of the form $e^{iL}$.  Note, however, that these perturbative contributions do not cancel in other quantities, for example $\langle Z(iT) \rangle$, the Fourier transform of the density.\footnote{See for example \cite{Morozov:2009uy}. This quantity has $1/N^q$ corrections in SYK which would be parametrically larger than these $1/L$ effects, possibly complicating their study.}  Such asymptotic series in $1/L^2$ typically grow like $(2g)!$ at genus $g$, compatible with $e^{iL}$ effects.   In matrix integrals these effects can be calculated using one eigenvalue instantons \cite{Neuberger:1980qh,Ginsparg:1990as,David:1990sk,Shenker:1990uf}.   In this case they are usually studied in the collective field formalism and called ``Altshuler-Andreev" instantons \cite{1995PhRvL..75..902A,1999JPhA...32.4373K}. 
  
 We now return to SYK.  In the low energy JT gravity regime the action \eqref{JTaction} contains a term proportional to the Euler character $\chi$ of the geometry and contributions  are weighted by $(e^{S_0})^{\chi}$.  But here $e^{S_0} \sim e^{N}$ so these are nonperturbative in $1/N$.   As discussed in previous sections the ramp is due to a saddle point with cylindrical topology.  No higher genus configuration contributes to the ramp, presumably because of the same sorts of cancellations that occur in GUE random matrix perturbation theory.  
 
 But there are indications that  higher genus configurations do play a role.   Correlation functions, discussed above, seem to be described by a handle attached to a disk ($\chi = -1$).   For the GOE ensemble (realized by SYK with certain $N$ values
  \cite{You:2016ldz})  the spectral form factor has an infinite series of $1/L$ corrections.\footnote{See for example \cite{mehta2004random}, Eq.~(7.2.46).  This expansion is convergent, presumably due to large but not exact cancellations.} These might well be due to nonorientable JT configurations.   But finding higher genus solutions to JT with the dilaton boundary conditions appropriate to SYK is challenging.\footnote{With zero dilaton, constant negative curvature metrics on higher genus Riemann surfaces are solutions.   The JT functional integral with these boundary conditions gives the Weil-Petersson  volume of moduli space, a problem famously connected to matrix models \cite{Dijkgraaf:2018vnm}. (We thank Edward Witten for pointing this out to us.)  This volume grows like $(2g)!$.} A nonzero solution of the dilaton equation implies the existence of a Killing vector \cite{Mann:1992yv}, absent at higher genus.  So some stabilizing effect that modifies the dilaton equation, or else a boundary condition that allows $\phi = 0$ as a solution would seem necessary.
 
 Nonetheless one  might conjecture that an asymptotic series of higher genus configurations exists.  These would define a kind of string theory, the JT string.  This theory would require a nonperturbative completion, including the analog of one eigenvalue instantons.   The $(2g)!$  perturbative behavior is generic in string theory \cite{Shenker:1990uf} where the corresponding nonperturbative effects are due to D-branes \cite{Polchinski:1994fq}.   So the plateau would be a D-brane effect in the JT string.

\section*{Acknowledgements}
We are grateful to Alex Altland, Hrant Gharibyan, Guy Gur-Ari, Daniel Jafferis, Alexei Kitaev, Juan Maldacena, Alex Maloney, David Gross, Leonard Susskind and Edward Witten for discussions, and to Raghu Mahajan for comments on the draft. PS and SS are supported in part by NSF grant PHY-1720397.  DS is supported by Simons Foundation grant 385600.

\appendix

\section{More details on Brownian SYK}\label{appendixa}

\subsection{Late-time ensembles}\label{app:latetimes}
It is interesting to consider the random matrix ensembles that the Brownian SYK models limit to as we take $t\rightarrow \infty$. In the case $q>2$ we expect the evolution to be generic enough to fill out the ensemble with the correct symmetry properties. There are two important discrete symmetries to consider \cite{Fidkowski:2009dba,You:2016ldz,Fu:2016yrv,Cotler:2016fpe,Kanazawa:2017dpd}
\begin{itemize}
\item The fermion parity symmetry $(-1)^F =(2i)^\frac{N}{2} \psi_1\psi_2\dots \psi_N$ commutes with the Hamiltonian, for any disorder realization. This implies that the Hamiltonians (and likewise the unitary we form by exponentiating them in the Brownian SYK evolution) are block diagonal, with two blocks corresponding to $(-1)^F = \pm 1$.
\item There is an antiunitary symmetry $\mathcal{T}$. To define the operator explicitly, we represent the $N$ Majorana fermions using $N/2$ spins with Pauli operators $X_i,Y_i,Z_i$:
\be
\psi_1 = \frac{1}{\sqrt{2}}X_1, \hspace{20pt} \psi_2 = \frac{1}{\sqrt{2}}Y_1, \hspace{20pt} \psi_3 = \frac{1}{\sqrt{2}}Z_1X_2, \hspace{20pt} \psi_4 = \frac{1}{\sqrt{2}}Z_1Y_2,\hspace{20pt}\dots
\ee
We work in the $Z$ basis so $\psi_a$ is real when $a$ is odd and imaginary when $a$ is even. $\mathcal{T}$ is defined as follows \cite{Fidkowski:2009dba}. For $N/2$ odd we write $\mathcal{T} = 2^{N/4}K \psi_1\psi_3\psi_5...\psi_{N-1}$. For $N/2$ even, we write $\mathcal{T} = 2^{N/4}K\psi_2\psi_4\psi_6...\psi_N$. Here $K$ is the antiunitary operator that takes the complex conjugate. One can check that $\mathcal{T}\psi_a\mathcal{T} = \psi_a$ and the following algebra
\begin{align}
N/2 &= 0 \text{ (mod 4): }\hspace{40pt} \mathcal{T}^2 = 1, \hspace{49pt} \mathcal{T}(-1)^F = (-1)^F\mathcal{T}\\
N/2 &= 1 \text{ (mod 4): }\hspace{40pt} \mathcal{T}^2 = 1, \hspace{49pt} \mathcal{T}(-1)^F = -(-1)^F\mathcal{T}\\
N/2 &= 2 \text{ (mod 4): }\hspace{40pt} \mathcal{T}^2 = -1, \hspace{40pt} \mathcal{T}(-1)^F = (-1)^F\mathcal{T}\\
N/2 &= 3 \text{ (mod 4): }\hspace{40pt} \mathcal{T}^2 = -1, \hspace{40pt} \mathcal{T}(-1)^F = -(-1)^F\mathcal{T}
\end{align}
\end{itemize}
Now we apply these considerations to the unitary constructed from the Brownian SYK evolution. First we consider the case where  $\mathbf{q = 0}$ {\bf (mod 4)}. In this case $\mathcal{T}$ commutes with a given realization of the Hamiltonian, and therefore anticommutes with $iH$. This means that
\be
\mathcal{T} \dots e^{-iH_3 \delta t}e^{-iH_2 \delta t}e^{-iH_1\delta t} \mathcal{T}^{-1} = \dots e^{iH_3\delta t}e^{iH_2\delta t}e^{iH_1\delta t}.
\ee
The operator on the RHS does not in general have any simple relationship to the operator on the LHS. So for the purposes of the Brownian SYK, we do not have any constraints from $\mathcal{T}$ symmetry. In this case the only symmetry restriction comes from the fact that the unitary $U(T)$ produced by the Brownian circuit commutes with $(-1)^F$. This means that $U(T)$ should be a direct sum of two blocks corresponding to even and odd fermion parity. We expect both of these blocks to approach independent elements $U_1,U_2$ of the Haar ensemble on the unitary group (CUE) as we take $T$ large, so that $\Tr[U(T)^k] \rightarrow \Tr[U_1^k] + \Tr[U_2^k]$, where the trace is over a $2^{N/2-1}$-dimensional Hilbert space of a single block. For the unitary Haar ensemble, one has
\be
\langle \Tr[U^k]\Tr[(U^*)^m]\rangle = \delta_{k,m}\text{min}(k,\text{dim}(U)).
\ee
So, for the Brownian SYK we expect at large $T$ to have
\be\label{expect1}
\langle |\Tr\,U(T)^k|^2\rangle \rightarrow 2\,\text{min}(k,2^{N/2-1}).
\ee

The case of $\mathbf{q =2}$ {\bf (mod 4)} is more interesting. Now $\mathcal{T}$ anticommutes with the Hamiltonian, because of the explicit factor of $i$ needed for Hermiticity, e.g. $H = i J_{a_1\dots a_6}\psi_{a_1}\dots\psi_{a_6}$. Note that this implies $\mathcal{T}$ commutes with $iH$, so we have
\be\label{q2}
\mathcal{T} \dots e^{-iH_3 \delta t}e^{-iH_2 \delta t}e^{-iH_1\delta t} \mathcal{T}^{-1} = \dots e^{-iH_3\delta t}e^{-iH_2\delta t}e^{-iH_1\delta t}.
\ee
i.e. $\mathcal{T}$ commutes with our unitary. We now have to consider several cases
\begin{enumerate}
\item If $N/2$ is odd, then $\mathcal{T}$ exchanges the two blocks corresponding to different values of $(-1)^F$. In this case the blocks are related by $\mathcal{T}$ and have complex-conjugate eigenvalues, but a single block has no symmetry constraints, and we expect that it will approach a Haar random unitary at late time. In other words, we expect the random matrix class CUE for one of the blocks. Then $\Tr[U(T)^k]\rightarrow \Tr[U^k] + \Tr[(U^*)^k]$ where $U$ is a random unitary. This leads, again, to (\ref{expect1}).
\item If $N/2$ is even then $\mathcal{T}$ preserves the blocks corresponding to a given value of $(-1)^F$. There are two further subcases:
\begin{enumerate}
\item If $N/2=0$ (mod 4), then  we have $\mathcal{T}^2 = 1$. By an argument described in \cite{mehta2004random} chapter 2, we can change basis so that $\mathcal{T}$ simply acts as $K$, complex conjugation. In this basis our unitary matrix must be real, in other words it should be an orthogonal matrix. We expect that at late times we will get a generic matrix from this ensemble, in other words (up to the change of basis) we expect the blocks to approach independent elements drawn from the Haar distribution on the orthogonal matrices with unit determinant. We will refer to this ensemble as $\widetilde{\text{CRE}}$, where the tilde indicates that we are restricting to the connected component of the identity (determinant one), because the Brownian evolution defines a path starting at the identity.

\item If $N/2=2$ (mod 4), then we have $\mathcal{T}^2=-1$. Then by a change of basis we can make the block a symplectic unitary matrix. We expect at late times to converge to the Haar distribution on $USp$. This is the class CQE.
\end{enumerate}
For these two cases we expect at late times that $\Tr[U(T)^k]\rightarrow \Tr[O_1^k] + \Tr[O_2^k]$ or $\Tr[U(T)^k]\rightarrow \Tr[S_1^k] + \Tr[S_2^k]$, where $O_1,O_2,S_1,S_2$ are elements of the orthogonal or symplectic groups. A new feature in these cases is that there is a ``disconnected'' piece in the spectral form factor, because $\langle\Tr[U(T)^k]\rangle$ by itself does not vanish for late time. Instead, based on formulas from \cite{pastur2004moments}, we expect
\be\label{onecopy}
\langle \Tr [U(T)^k]\rangle \rightarrow \pm (1 + (-1)^k)
\ee
where the positive sign is for case (a) and the negative sign for case (b). One also finds
\be\label{q2mod4}
\langle |\Tr\,U(T)^k|^2\rangle \rightarrow 2k + (1 + (-1)^k)^2.
\ee
The additional term is the square of (\ref{onecopy}), so the {\it connected} spectral form factor is the same as in previous cases. These formulas are expected to be valid provided that $k$ is less than $2^{N/2-2}$. See \cite{pastur2004moments} for formulas that are accurate also for larger values of $k$.

\end{enumerate}

\subsection{The \texorpdfstring{$N$}{N} mod 8 periodicity from saddle points}\label{app:Nmod8}
It is interesting to see how the detailed structure described in the previous section can arise from saddle points in the Brownian SYK model. Since the subtlety is in the disconnected term in the spectral form factor, we will focus on $\langle \Tr[U(T)^k]\rangle$. For this problem we use $k$ replicas of the fermions, $\psi^{(s)}_a$, where $a = 1,...,N$ is the usual flavor index and $s = 1,...,k$ is the replica index. The correct boundary conditions to compute the quantity we are interested in are the ones in (\ref{bckreplica}).

To write a $G,\Sigma$ action, we need to use matrices $G^{s,s'}(t),\Sigma^{s,s'}(t)$ where antisymmetry allows us to restrict to $s>s'$. The analog of the path integral (\ref{GSigma}) is
\begin{align}
\langle \Tr[U(T)^k]&\rangle \approx \int \mathcal{D}G^{s,s'}\mathcal{D}\Sigma^{s,s'} \exp\left\{-\frac{N}{2}\int_0^T dt \left[\frac{2J}{q}\left(\frac{k}{2^q} +\sum_{s>s'}(G^{s,s'})^q\right) + \sum_{s>s'}\Sigma^{s,s'}(t)G^{s,s'}(t)\right]\right\}\notag\\
&\hspace{10pt}\times \int \mathcal{D}\psi_a^{(s)}\exp\left\{-\frac{1}{2}\int_0^T dt \left[\psi_a^{(s)}\partial_{t}\psi_a^{(s)}   - \sum_{s>s'}\psi_a^{(s)}(t)\psi_a^{(s')}(t)\Sigma^{s,s'}(t)\right]\right\}.\hspace{-20pt}\label{GSigmak}
\end{align}
For large $T$ and even $k$, this integral has saddle points consistent with the boundary conditions (\ref{bckreplica}) where we correlate the fermions with their partners ``halfway around the circle:''
\be\label{halfway}
G^{s,s+\frac{k}{2}}(t) = \pm \frac{i}{2}, \hspace{20pt} \Sigma^{s,s+\frac{k}{2}}(t) = \mp\frac{iJ}{2^{q-2}}
\ee
and we set all other $G,\Sigma$ components to zero. The notation here requires some explanation. In cases where the index $s+k/2$ is larger than $k$, it should be interpreted in the cyclic sense, including the minus sign from the fermion antiperiodicity, so that e.g. $G^{s,k+1} = -G^{s,1}$. We also are working with the understanding that $G^{ss'} \rightarrow -G^{s's}$ in cases where $s'>s$, and similarly for $\Sigma$.

An important point is that for this saddle point when $q = 2$ (mod 4), the $G^q$ term on the first line of (\ref{GSigmak}) cancels the $\frac{k}{2^q}$ arising from the $\psi(t+\epsilon)\psi(t) = \frac{1}{2}$ terms. This is the key feature that allows these saddle points to contribute at late times when $q = 2$ (mod 4) but not otherwise.

To evaluate the fermion determinant, we can combine the $k$ segments together into one antiperiodic circle. The path integral we want is $N$ copies of an integral of the form
\be
\int \mathcal{D}\psi \exp\left\{-\frac{1}{2}\int_0^{kT}\psi(t)\partial_t\psi(t) - \psi(t)\psi(t+\frac{k}{2}T)\Sigma\right\}.
\ee
One way to evaluate this is to decompose into modes
\be
\psi(t) = \sum_n e^{i\omega_n t}\psi_n, \hspace{20pt} \omega_n = \frac{2\pi(n+\frac{1}{2})}{kT}.
\ee
The separate integrals over the different modes then give us a product representation for the answer. To regularize the product, we can use that in the case with $\Sigma = 0$ we should get $\sqrt{2}$. This leads to
\be\label{factor}
\sqrt{2}\prod_{m = 0}^\infty \left(1 - \frac{(-1)^m \Sigma}{\omega_m}\right) = \sqrt{2}\left(\cos\frac{k\Sigma T}{4} - \sin\frac{k\Sigma T}{4}\right).
\ee
Using this formula, we can now evaluate the full integrand in (\ref{GSigmak}) for the configuration (\ref{halfway}). For large $T$ and assuming $q = 2$ (mod 4), the answer is simply $e^{\pm \frac{2\pi i}{8}N}$, where the upper and lower signs correspond to the ones in (\ref{halfway}). This phase comes entirely from the phase of the factor (\ref{factor}), which for pure imaginary $\Sigma$ is proportional to $\frac{1\pm i}{\sqrt{2}} = e^{\pm\frac{2\pi i}{8}}$. 

This phase has an obvious $N$ mod 8 periodicity. Summing over the two signs, we find that the two saddles cancel when $N/2$ is odd, add to something positive when $N/2 = 0$ (mod 4), and add to something negative when $N/2 = 2$ (mod 4). This is consistent with the pattern identified in the previous section.

\subsection{Exact evaluation of \texorpdfstring{$\langle |\Tr\,U(T)|^2\rangle$}{<|Tr U(T)|2>}}\label{exactapp}
Although it is not necessary for our discussion, it is straightforward to get an exact expression for $\langle |\Tr\,U(T)|^2\rangle$. This is easiest using the auxiliary spin Hamiltonian (\ref{spintheory}). The eigenstates of $H_{spin}$ are states in which $m$ of the spins are up, and the rest are down. The corresponding eigenvalue can be calculated as
\be
E_{spin}(m) = \hat{J}\binom{N}{q} - \hat{J}\sum_{k = 0}^{\text{min}(m,q)} \binom{N-m}{q-k}\binom{m}{k}(-1)^k, \hspace{20pt} \hat{J} \equiv \frac{J}{2^q\binom{N}{q-1}}.
\ee
Taking into account the multiplicity $\binom{N}{m}$, we find
\be
\langle |\Tr\,U(T)|^2\rangle = \Tr[e^{-T H_{spin}}] = \sum_{m = 0}^N\binom{N}{m}e^{-E_{spin}(m) T}.
\ee

Before we try to obtain this exact result in the $G,\Sigma$ variables, it is helpful to write it in a different way. We can rewrite the Hamiltonian in terms of a combined spin variable $S_z = \sum_a \sigma^{(z)}_a$ using
\be
q!\sum_{a_1<...<a_q}\sigma_{a_1}^{(z)}\dots\sigma_{a_q}^{(z)} = \sum_{a_i\text{ distinct}}\sigma_{a_1}^{(z)}\dots\sigma_{a_q}^{(z)} = f_q(S_z).
\ee
Although the exact function $f_q(x)$ will not be necessary, one can get a recurrence relation for it by considering
\begin{align}
S_z f_q(S_z) &= \sum_b \sigma_b^{(z)} \sum_{a_i\text{ distinct}}\sigma_{a_1}^{(z)}\dots\sigma_{a_{q}}^{(z)}= f_{q+1}(S_z) + q(N{+}1{-}q)f_{q-1}(S_z).
\end{align}
Here the first term reflects the case where $b \neq a_i$ and the second term reflects the case where $b$ is equal to one of the $a_i$, and we used $(\sigma_b^{(z)})^2=1$. This recurrence relation, together with the initial conditions $f_0 = 1,$ $f_1(x) = x$ determines $f_q(x)$. We can now use this to write an alternate exact formula
\begin{align}
\Tr[e^{-TH_{spin}}] &= \Tr \left[  e^{-\hat{J}T\left[\binom{N}{q} - \frac{1}{q!}f_q(S_z)\right]}\right]= \int \frac{dy dx}{2\pi}\Tr\left[e^{i(x-S_z)y}\right]e^{-\hat{J}T\left[\binom{N}{q}-f_q(x)\right]}\\
&=\int \frac{dxdy}{2\pi} (2\cos y)^Ne^{ixy-\hat{J}T\left[\binom{N}{q}-f_q(x)\right]}.\label{sameas}
\end{align}

It is interesting to see how to arrive at this same formula from the $G,\Sigma$ approach. As a first step we need to write a more precise formula for the action of the $G,\Sigma$ theory. There are two imprecisions with what we wrote in (\ref{GSigma}). The first is that we used large $N$ approximations to the binomial coefficients. The second is that we represented the fermion interaction term as $G(t)^q$. This is imprecise because $G^q$ contains terms where we have some indices the same. This is the same problem that we encountered in the last paragraph, and we can fix it with the same function $f_q(x)$, replacing
\be
i^qG_{LR}^q(t)\rightarrow \frac{1}{(2N)^q}f_q\left(2iNG_{LR}(t)\right). 
\ee 
These improvements give an exact version of the action in (\ref{GSigma}), but we still have to evaluate the path integral. A key point is that the fermion determinant depends only on the average value of $\Sigma_{LR}(t)$. To show this, one can recognize the second line of (\ref{GSigma}) as the partition function of a theory with time-dependent Hamiltonian $H(t) = -\frac{\Sigma_{LR}(t)}{2}\sum_a\psi_a^{(L)}\psi_a^{(R)}$. Since the Hamiltonian commutes with itself at different times, only the time-integral of $\Sigma_{LR}(t)$ matters. 
When we integrate over modes of $\Sigma_{LR}$ with zero time average, the only dependence is in the first line of (\ref{GSigma}), and these variables act as Lagrange multipliers setting the corresponding non-constant modes of $G_{LR}(t)$ to zero. We find that the entire integral reduces to an integral over the constant values of $G_{LR}$ and $\Sigma_{LR}$:
\be
\langle |\Tr\,U(T)|^2\rangle = \int dG_{LR} d\Sigma_{LR} \left(2\cos\frac{T\Sigma_{LR}}{4}\right)^Ne^{-\frac{NT}{2}G_{LR}\Sigma_{LR}- \hat{J}T\left[\binom{N}{q} - \frac{1}{q!}f_q(2iNG_{LR})\right]}.
\ee
After changing variables to $x = 2iNG_{LR}$ and $y = \frac{T\Sigma_{LR}}{4}$, and normalizing the integration measure so that we get $2^N$ when $J = 0$, we recover the same integral expression as (\ref{sameas}).

\subsection{Weingarten functions and finite \texorpdfstring{$L$}{L} effects}\label{finiteL}
In Section \ref{zk} and Appendix \ref{app:Nmod8}, we sketched how saddle points can account for behavior of $\langle |\Tr[U(T)^k]|^2\rangle$ that is expected based on the random matrix ensembles described in Appendix \ref{app:latetimes}. More precisely, the saddle points seem to explain the behavior for $k \le 2^{N/2 -2}$, but the random matrix ensembles suggest that the answer should change qualitatively for $k > 2^{N/2 -2}$ (or $2^{N/2 -1}$, depending on $N$ and $q$) indicating that there are other important contributions to the path integral.  This is a nonperturbative effect in $1/L$ (where we write  $L = 2^{N/2}$ for the dimension of the full Hilbert space) in the sense that the small $k$ behavior is exactly independent of $L$ for $k < L/4$. However, in somewhat different quantities one expects perturbative $1/L$ effects. 

Here we discuss this a bit further, emphasizing corrections that are perturbative in $1/L$ and can in principle be calculated in the $G, \Sigma$ description of the SYK Brownian circuit. Our discussion builds on the standard approach to Haar integrals over unitary matrices described in \cite{Collins:550493,2006CMaPh.264..773C,Roberts:2016hpo}. We begin by considering $2k$ copies of the Hilbert space, where $L_1,...,L_k$ will be acted on by $U$, and $R_1,...,R_k$ will be acted on by $U^*$.  We define $A$ to be the Haar average of the tensor product of $2k$ unitary operators, acting on such a Hilbert space:
\be
A= \int dU\,\underbrace{U\otimes ...\otimes U}_{\text{acting on }L_1...L_k}\otimes\underbrace{ U^*\otimes...\otimes U^*}_{\text{acting on }R_1...R_k}.
\ee
A useful fact is that $A$ is equal to the projector onto the subspace spanned by $k!$ maximally entangled states. These states are defined by pairing up and maximally entangling $L_1,...,L_k$ with $R_1,...,R_k$ according to some permutation $\sigma \in S_k$:
\be\label{maxstates}
|\sigma\rangle = |\text{MAX}\rangle_{L_1,R_{\sigma(1)}}\otimes ...\otimes |\text{MAX}\rangle_{L_k,R_{\sigma(k)}}, \hspace{20pt} |\text{MAX}\rangle_{AB} \equiv \frac{1}{L^{1/2}}\sum_{i = 1}^L|i\rangle_A\otimes |i\rangle_B.
\ee
To prove this, one can start with the usual formula for the Haar average
\be\label{momentsofU}
\int dU\, U_{i_{1}j_{1}}\cdots U_{i_{k}j_{k}}U_{i_{1}^{\prime }j_{1}^{\prime }}^{*}\cdots U_{i_{k}^{\prime }j_{k}^{\prime }}^{*} =  \sum _{\sigma ,\tau}\delta _{i_{1}i_{\sigma(1)}^{\prime }}\cdots \delta _{i_{q}i_{\sigma(k)}^{\prime }}\delta _{j_{1}j_{\tau(1)}^{\prime }}\cdots \delta _{j_{k}j_{\tau(k)}^{\prime }}W_{\sigma, \tau}
\ee
where $\sigma,\tau$ are permutations and the coefficients $W_{\sigma,\tau}$ are known as the Weingarten functions. (\ref{momentsofU}) can be rewritten as
\be\label{projeqn} 
 A = L^k\sum_{\sigma,\tau}  W_{\sigma,\tau} |\sigma\rangle\langle\tau|.
\ee
Now, it is easy to check that $U\otimes U^*|\text{MAX}\rangle = |\text{MAX}\rangle$ for any unitary $U$, which implies that $A|\sigma\rangle = |\sigma\rangle$. Because of (\ref{projeqn}), this implies $A^2 = A$. So $A$ is a projection operator that preserves the states (\ref{maxstates}) and acts within their span. This means that $A$ is the projection operator onto this subspace.

The finite $L$ effects (both perturbative and nonperturbative) are due to the fact that these states are not orthonormal, with overlap
\be \label{maxoverlap}
\langle\tau|\sigma\rangle=L^{c(\tau^{-1}\sigma)-k}
\ee
where $c(\tau^{-1}\sigma)$ is the number of cycles in the permutation $\tau^{-1} \sigma$. The Weingarten coefficients are roughly the matrix inverse of this inner product, $\langle \tau|\sigma\rangle W_{\sigma,\rho} = L^{-k}\delta_{\tau,\rho}$, which is invertible for large enough $L$. This relation can be used to determine $W_{\sigma,\rho}$.

At infinite $L$, the states are orthogonal, and in particular they are linearly independent. This implies that $\Tr A = k!$ in this limit. We can try to understand the analogous behavior in the SYK Brownian circuit. For simplicity we consider the case where $q = 0$ (mod 4). In appendix \ref{app:latetimes}, we argued that $\Tr[U(T)]\rightarrow \Tr[U_1] + \Tr[U_2]$, where $U_1,U_2$ are independent Haar random unitaries. Then $\Tr[A] = |\Tr[U(T)]|^{2k} = |\Tr U_1 + \Tr U_2|^{2k}$. Averaging separately over $U_1,U_2$, one finds $2^k k!$. Now, we can represent $|\Tr[U(T)]|^{2k}$ as a $G, \Sigma$ integral as in \ref{zk} with fields $G_{ij}^{s s'}(t), \Sigma_{ij}^{s s'}(t)$, where $s, s' = 1 \ldots k$ and $i,j$ are $L$ or $R$.  Because each of the $2k$ sectors are independent there is no boundary condition linking different segments. There is a natural set of saddle points, as in \ref{zk}, where the $s$ and $s'$ indices are paired according to one of the $k!$ permutations in $S_k$.   For this set of paired indices $G_{LR}^{s s'}, \Sigma_{LR}^{s s'}$ are set equal to one of the two values in \eqref{nontrivialsaddle}.  Each of these has action $0$ at large $T$ so the number of saddle points is $2^k k!$.\footnote{In the case $q = 2$ (mod 4), there is no distinction between the $L$ and $R$ systems, so we can form more saddle points, where we pair up the $2k$ quantities in any fashion, and assign $\pm$ signs to each pair. This leads to $2^k(2k-1)!!$ saddle points. Using \cite{pastur2004moments}, one can check that this is the same answer we get for each of the $q = 2$ (mod 4) ensembles in appendix \ref{app:latetimes}.} We have not shown that each saddle contributes precisely one to the integral, but it seems clear that the saddle points are capable of giving the correct answer for large $L$.

Now we discuss finite $L$. In the $1/L$ expansion, the states (\ref{maxstates}) acquire nontrivial overlaps, but for sufficiently large $L$ they remain linearly independent, so $\Tr A = k!$ remains exactly correct, with deviations that are not visible in the $1/L$ expansion (they start at $L = k-1$, see below). We are not certain how this manifests itself in the $G,\Sigma$ description. An easier target might be to look at quantities for which there are perturbative $1/L$ corrections.

As an example consider the two $k=2$  Weingarten coefficients:
\begin{eqnarray}\label{kequaltwow}
W_{\sigma,\tau} = \begin{cases} 
      \frac{1}{L^2-1} & \tau^{-1}\sigma = \text{identity permutation} \\
      \frac{-1}{L(L^2 -1)} & \tau^{-1}\sigma = \text{swap}.
   \end{cases}
\end{eqnarray}
These functions have an infinite series of $1/L$ corrections.  The poles in these functions indicate that the matrix \eqref{maxoverlap} is no longer invertible for $L \le 1$.  For general $k$ there are analogous poles indicating the lack of invertibility (which is the same as linear dependence of the states (\ref{maxstates})) for small enough $L$ relative to $k$.   The dimension of the full Hilbert space is $L^{2k}$ and the number of maximally entangled states is $k!$ so if the maximally entangled states were generic vectors in the Hilbert space one would expect problems when $k! \sim L^{2k}$ or, roughly,  $k \sim L^2$.   In fact \eqref{maxoverlap} shows that the maximally entangled states are not generic.  It is known \cite{Collins:550493} that the Weingarten functions have simple poles for $L = -(k-1), -(k-2) \ldots (k-2), (k-1)$.  So the first place where there is noninvertibility is $k = L+1$.\footnote{Using random matrix techniques one can show for large $k, L$ and $k \sim L^2$ that the dimension of the maximally entangled space is $k!$ up to exponentially small corrections for $k < k_c \sim L^2$, the location of a Gross-Witten-Wadia large $N$ transition.}   

It is easy to find quantities that are sensitive to these $1/L$ corrections. For example,
\be 
\langle \Tr [U^2 (U^*)^2] \rangle = \frac{L+2}{L+1} = 1 -\sum_{n=1}^{\infty} \left(\frac{-1}{L}\right)^n.
\ee
This quantity has a $G, \Sigma$ description for the SYK Brownian circuit.  Since $1/L \sim e^{-c N}$  the $1/L$ corrections should come from nonperturbative effects in the  $G, \Sigma$ dynamics. Perhaps a more direct route would be to directly compute the analog of the overlap \eqref{maxoverlap}. The $G, \Sigma$ saddle points discussed above correspond in a natural way, as in the $k=1$ case, to certain fermion states.  In the $k=1$ case these are just the maximally entangled states.  For $k >1$ they are linear combinations of the maximally entangled states \eqref{maxstates}.  At large $L$ these states are orthonormal, but the almost orthogonal states have an overlap that is a power of $1/L$.  We hope that an  understanding of the collective field origin of these effects will shed some light on how the finite dimensionality of the Hilbert space is coded into this description.  In particular we hope that this calculation will shed some light on the $k \gg L$ limit.

In closing we mention that the double cone provides a calculation of an overlap of almost orthogonal states.   Let $|{\rm TFD}\rangle$ be the properly normalized thermofield double state.  Denote the time shifted state $e^{-i H_R T}|{\rm TFD}\rangle$ by $|{\rm TFD}(T)\rangle.$  Then as Papadodimas and Raju \cite{Papadodimas:2015xma} pointed out 
\be\label{tfdover}
|\langle {\rm TFD}| {\rm TFD}(T)\rangle|^2 ~= \frac{1}{Z(\beta)^2} Z(\beta+iT)Z(\beta-iT) ~.
\ee

Two randomly chosen vectors $|v\rangle, |w\rangle$ in a Hilbert space of dimension $d$ have squared overlap  $|\langle v | w \rangle|^2 \sim \frac{1}{d}$.  The dimension of the doubled Hilbert space is $L^2$.   Working at $\beta =0$  the size of the early part of the ramp in \eqref{tfdover} is of order $1/L^2$, indicating uncorrelated states.\footnote{For simplicity we ignore the slope contribution.}   But at later time the overlap increases, eventually saturating at order $1/L$ on the plateau.  This reflects the fact that $e^{-i H_R T}$ does not approach a random unitary at late time, as we now discuss.

\subsection{Why the plateau of Brownian SYK is \texorpdfstring{$O(1)$}{O(1)} rather than \texorpdfstring{$O(L)$}{O(L)}}
In Brownian SYK the quantity $\langle |\Tr\,U(T)|^2\rangle$ approaches an order one value at late times, while in regular SYK the analogous $\langle |Z(iT)|^2\rangle$ approaches an exponentially large value, $2^{\frac{N}{2}}$ or $2^{\frac{N}{2}+1}$. This difference is reflected in a qualitative difference between two types of similar-sounding ensembles of unitary matrices: CUE and an ensemble we'll call GUE${}_\infty$.

CUE is simply the invariant measure on the group $U(L)$. It defines a good notion of a totally random unitary. GUE${}_\infty$ is defined by taking a random Hamiltonian from the GUE ensemble, and running the time evolution for a very long time. More explicitly,
\be
\langle F(U)\rangle_{\text{CUE}} = \int_{Haar} dU\,F(U), \hspace{40pt} \langle F(U)\rangle_{\text{GUE}_{\infty}} = \lim_{T\rightarrow \infty} \int dH\,e^{-L\Tr(H^2)} F(e^{-iHT}).
\ee

The statistics of the eigenvalues in the two ensembles are very different, as emphasized to us by Michael Berry \cite{Berry}. In CUE, the eigenvalues of $U$ are phases $e^{i\theta_k}$, and the angles $\theta_k$ exhibit level repulsion, spectral rigidity, and in general have interesting statistical properties. For example
\begin{align}
\langle |\Tr(U^k)|^2\rangle_{\text{CUE}} &= L + \sum_{n\neq m}\langle e^{ik(\theta_n-\theta_m)}\rangle_{\text{CUE}}=L + \int \frac{d\theta d\theta'}{(2\pi)^2}e^{ik(\theta-\theta')}\left[L^2 - \frac{\sin^2(L\frac{\theta-\theta'}{2})}{\sin^2(\frac{\theta-\theta'}{2})}\right] 
\end{align}
which evaluates to $\text{min}(k,L)$. On the other hand, in GUE, the eigenvalues are $e^{iE_jT}$. The quantities $E_j$ have similar statistical properties to the $\theta_j$, but these get washed out after multiplying by a large factor $T$ and evaluating mod $2\pi$. So in GUE${}_\infty$, the eigenvalues of the unitary are completely uncorrelated, and we have
\be
\langle |\Tr(U^k)|^2\rangle_{\text{GUE}_\infty} = L + \sum_{n\neq m}\langle e^{ik(\theta_n-\theta_m)}\rangle_{\text{GUE}_\infty} = L + L(L{-}1)\int\frac{d\theta d\theta'}{(2\pi)^2}e^{ik(\theta-\theta')}  = L.
\ee

The difference between the ensembles can also be stated like this: the GUE${}_\infty$ ensemble amounts to starting at the identity in $U(L)$, picking a random direction in the tangent space, and traveling along it for a very long time. This does not lead to a uniform distribution on $U(L)$. (An analogous statement for $S^2$ is easy to visualize.)

\section{Showing the action is zero for the SYK saddle points}\label{actionapp}
In this appendix we show that the action of the SYK saddle points vanishes for large $T$. It's convenient to rescale time by a factor of $T$ so that it runs $0<t<1$. Then the full action, including the $\beta$ variables is
\begin{align}
NI_{full} &= -\left[ (\beta_L+\beta_R) E + (\beta_L^2+\beta_R^2)\DeltaE^2\right]-N\text{log Pf}\left(\delta_{ij}\partial_t -\Sigma_{ij}\right) \notag\\&\hspace{20pt}+ \frac{N}{2}\int_0^1\int_0^1 dt dt'\left[ \Sigma_{ij}G_{ij} - \frac{\widetilde{J}_i\widetilde{J}_j}{q}s_{ij}G_{ij}^q\right].\label{actionfullapp}
\end{align}
where we compensate for the rescaled time by rescaling the couplings
\be
\widetilde{J}_L = (T - i\beta_L)J, \hspace{20pt} \widetilde{J}_R = (T+i\beta_R)J.
\ee
We will argue that the action vanishes in two steps. First, we show that it is independent of time for large $T$. To do this, we vary the on-shell action with respect to $T$. Because the action is stationary with respect to small changes in $G,\Sigma$, we only need to vary the explicit factors of $T$. For large $T$ the solutions we considered have $\beta_L = \beta_R = 0$, and in this case one finds
\be
\partial_T (N I_{full}) = NTJ^2\int_0^1\int_0^1 dt dt' \left[G_{LL}^q+G_{RR}^q - i^q(G_{LR}^q + G_{RL}^q)\right].
\ee
This is proportional to $i(E - E) = 0$ by (\ref{energyconstraint}), so the time derivative of the action vanishes. In fact, the vanishing of the RHS also shows that the $s_{ij}G_{ij}^q$ term in (\ref{actionfullapp}) and (after using the equation $\Sigma_{ij} \propto s_{ij}G_{ij}^{q-1}$) the $\Sigma_{ij}G_{ij}$ term are zero.

To analyze the remaining $\log \text{Pf}$ term, it is convenient to rescale time back again so that $0<t<T$. In Fourier space, the regularized Pfaffian is
\be
\log\text{Pf}(\delta_{ij}\partial_t - \Sigma_{ij}) \rightarrow \log(2) + \frac{1}{2}\sum_{n=-\infty}^\infty\log \det \left(\begin{array}{cc}1 + \frac{\Sigma_{LL}(\omega_n)}{i\omega_n} & \frac{\Sigma_{LR}(\omega_n)}{i\omega_n}\\ \frac{\Sigma_{RL}(\omega_n)}{i\omega_n} & 1 + \frac{\Sigma_{RR}(\omega_n)}{i\omega_n}\end{array}\right)\label{tocancel}
\ee
Here we have subtracted the free determinant and added $\log(2)$ so that we get $2^N= Z(0)^2$ in the free theory. For large $T$ the Matsubara frequencies $\omega_n = \frac{2\pi(n+1/2)}{T}$ are closely spaced, and naively we can approximate the sum by an integral over continuous frequency. In order to do this, $\Sigma_{ij}(\omega)$ can be defined for continuous $\omega$ following the steps in (\ref{logicin}). This gives a function of continuous $\omega$ that agrees with  $\Sigma(\omega_n)$ at the Matsubara frequencies up to exponentially small (in $JT$) corrections. One would like to apply the Poisson resummation formula, which for our case reads
\be
\sum_{n=-\infty}^\infty f(\omega_n) = \frac{T}{2\pi}\sum_{k = -\infty}^\infty (-1)^k\int_{-\infty}^\infty d\omega  e^{-i T k \omega} f(\omega).
\ee
For $f(\omega)$ analytic in a strip surrounding the real axis, the $k\neq 0$ terms are exponentially small in $T$, and the $k = 0$ term is proportional to $T$. 

Now, because $\Sigma$ is exponentially decaying in real time, $\Sigma_{ij}(\omega)$ is analytic in a strip surrounding the real axis. However, because of the factors of $1/\omega_n$, the continuation of (\ref{tocancel}) has a singularity at $\omega = 0$. For small $\omega$ we have that $\Sigma_{LL}(\omega)$ and $\Sigma_{RR}(\omega)$ are $O(\omega)$ for small $\omega$, but $\Sigma_{LR}(\omega) = -\Sigma_{RL}(\omega)$ approaches an imaginary constant $i\sigma$ plus $O(\omega^2)$. If we subtract $\frac{1}{2}\log (1 + \frac{\sigma^2}{\omega_n^2})$ from (\ref{tocancel}), we remove the singularity and can use the Poisson summation formula to conclude the answer is linear in $T$ plus exponentially small in $T$. However, we need to add back the term we subtracted, which can be evaluated explicitly
\be
\frac{1}{2}\sum_{n=-\infty}^\infty\log(1 + \frac{\sigma^2}{\omega_n^2}) = \log\cosh(\frac{\sigma T}{2}) = \frac{\sigma T}{2} -\log(2) + O(e^{-\sigma T}).
\ee
The $\log(2)$ cancels the one in (\ref{tocancel}), so we conclude that the $\log \text{Pf}$ term is linear in $T$, plus terms that are exponentially small in $T$. The term linear in $T$ (plus the $k = 0$ term from the Poisson resummation) must in fact be zero by the first step in our argument. So the entire action is exponentially small in $T$ for large $T$.

\section{The one-loop determinant (new in v2)}\label{app:oneloop}
\subsection{Setting up the determinant}
In this appendix we will compute the one-loop determinant about the saddle points discussed in section \ref{secbeta0}. In general in SYK, the one-loop determinant about a saddle point of the $G,\Sigma$ collective field action is given by a sum over the eigenvalues of the ladder kernel \cite{polchinskiStreicher},
\be
\log(\text{1-loop det.}) = -\frac{1}{2}\sum_{\text{eigenvalues }k}\log(1 - k).
\ee
The ladder kernel is constructed from the $G$ configuration that defines the saddle point. We are interested in evaluating the determinant about the saddle points discussed in section \ref{secbeta0}, where we have two replicas and the saddle point correlators $G_{ij}(t,t')$ where $i,j = L,R$. Explicitly, for this case the (symmetrized) ladder kernel is
\be
\hat{K}(1,2;3,4) = (q-1)J^2 \widetilde{G}(1,2)^{\frac{q-2}{2}}G(1,3)G(2,4)\widetilde{G}(3,4)^{\frac{q-2}{2}}.
\ee 
Here we used a combined notation for time and replica arguments, e.g. $G(1,2) = G_{i_1i_2}(t_1,t_2)$. We also introduced tilde-ed quantities\footnote{The formula for the regular (unsymmetrized) kernel for our two-replica Lorentzian problem is
\be
K(1,2;3,4) = -(q{-}1)J^2G(1,3)G(2,4)G(3,4)^{q-2}s_{i_3i_4} = (q{-}1)J^2G(1,3)G(2,4)\widetilde{G}(3,4)^{q-2}
\ee
and the symmetrized kernel is obtained by conjugating by $\widetilde{G}^\frac{q-2}{2}$.}
\begin{align}
\widetilde{G}_{LL}(t,t') &= \text{sgn}(t{-}t')G_{LL}(t,t'), \hspace{20pt} \widetilde{G}_{RR}(t,t') = \text{sgn}(t{-}t')G_{RR}(t,t')\notag\\
\widetilde{G}_{LR}(t,t') &= -iG_{LR}(t,t'), \hspace{54pt} \widetilde{G}_{RL}(t,t') = -iG_{RL}(t,t').\label{widetildeG}
\end{align}
The kernel should be viewed as an operator acting on functions of the form $f(3,4) = f_{i_3i_4}(t_3,t_4)$. For numerical computations we discretize this space; then the kernel becomes a finite dimesional matrix.

Because the ramp saddle points discussed in this paper preserve a time translation symmetry (acting together on the $L,R$ systems), the kernel can be partially diagonalized by going to frequency space for the sum of the two time arguments, and writing eigenvectors as
\be\label{partialDiag}
f_{i_3i_4}(t_3,t_4) = e^{\frac{i}{2}\omega_n(t_3+t_4)}\,\widetilde{f}_{i_3i_4}(t_{34}), \hspace{20pt} \omega_n \equiv \frac{2\pi n}{T}, \hspace{20pt} t_{34} \equiv t_3-t_4.
\ee
 For each Matsubara frequency $\omega_n$ there are still an infinite set of eigenvectors corresponding to different choices of $\widetilde{f}$. These can be labeled by a second abstract index $h$, so that all together the eigenvalues are denoted $k_T(\omega_n,h)$:
 \be
\log(\text{1-loop det.}) = -\frac{1}{2}\sum_{n,h}\log[1 - k_T(\omega_n,h)].
 \ee

One simplifying feature is that we are interested in evaluating the determinant assuming that $TJ \gg 1$, so that the exponential decay of the real-time correlators is very rapid compared to the periodicity $T$. In this case, the correlators $G$ for fixed time arguments approach $T$-independent values. This means that as we increase $T$, we enlarge the matrix corresponding to the kernel, but (apart from the volume of a time translation zero mode that determines the spectrum of Matsubara frequencies) we don't substantially change the existing matrix elements. The new matrix elements that get added correspond to one or more of the times $t_{12}$ and $t_{34}$ being roughly $T/2$. Because of the factors of $G(1,2)^{\frac{q-2}{2}}$ and $G(3,4)^\frac{q-2}{2}$, the new matrix elements we are adding are exponentially small. The partial sums over $h$ should therefore approach a fixed function of $\omega$:
\be
-\frac{1}{2}\sum_h \log[1 - k_T(\omega_n,h)] = g(\omega_n) + O(e^{-(\text{const})T})
\ee
where $g(\omega)$ is a function of continuous frequencies, defined in the infinite $T$ problem:
\be
g(\omega) = -\frac{1}{2}\sum_h \log[1 - k_{\infty}(\omega,h)].
\ee

Let's temporarily make a naive assumption that $g(\omega)$ is analytic in a strip of some width surrounding the real axis. Then the determinant we want could be written using the Poisson summation formula as
\be\label{poisson}
\log(\text{1-loop det.}) \stackrel{?}{=} \frac{T}{2\pi}\sum_{m = -\infty}^\infty\int_{-\infty}^\infty d\omega \, e^{imT\omega} g(\omega) \stackrel{?}{=} \# T + O(e^{-\# T})
\ee
where the linear term on the RHS comes from the $m = 0$ term in the sum, and the other terms are exponentially small because by our naive assumption $g(\omega)$ is analytic enough to allow the contour to be deformed slightly above or below the real axis. The linear term would have to vanish. This is because it is a local term that would also be present for the quantity
\be\label{indoftime}
\text{Tr}[e^{-bH}e^{iHT}e^{-bH}e^{-iHT}],
\ee
for which the $G,\Sigma$ saddle point and kernel along the long Lorentzian portions of the contour are the same as for the present problem, see figure \ref{figtwocontours}. If the linear term in (\ref{poisson}) were present, it would give exponential time dependence to (\ref{indoftime}). But (\ref{indoftime}) is exactly independent of $T$ so the linear term must vanish. Crucially, (\ref{poisson})  has no constant term as a function of $T$, so the vanishing of the linear term implies that for large $T$ the determinant is simply one.

Mathematically, the problem with this argument is that $g(\omega)$ is not actually analytic in a strip surrounding the real axis. This is due to the existence of a pair of zero modes with eigenvalues of the kernel $k = 1$. These lead to a singularity at $\omega = 0$. As discussed in section \ref{secbeta0}, these zero modes correspond to {\it (i)} a relative shift $\Deltat$ of time coordinates on the $L,R$ systems, and {\it (ii)} changes to the arbitrary auxiliary temperature $\beta_{\text{aux}}$ used to construct the saddle point. In addition to these exact zero modes, there are two families of soft modes that we get by making $\Deltat$ and $\beta_{\text{aux}}$ vary slowly in time, $\Deltat(t)$ and $\db_{\text{aux}}(t)$. If we remove these two functional degrees of freedom from the one-loop determinant, then the argument in the previous paragraph is valid, and it allows us to ignore the contribution from all of the ``generic'' modes. We will need to study the integral over the soft modes $\Deltat(t)$ and $\db_{\text{aux}}(t)$ explicitly, but this is a significant reduction in complexity.

\subsection{The hydrodynamic modes}
The two soft modes $\Delta(t),\db_{\text{aux}}(t)$ can be understood as arising from hydrodynamic fluctuations of the two systems $L$ and $R$. To describe them, we need an effective theory for these fluctuations. Fortunately, effective field theory for hydrodynamics has been studied extensively in the recent literature, starting with \cite{Dubovsky:2005xd}. The particular problem we are interested in, with two systems $L$ and $R$, is very similar to the problem of formulating hydrodynamics on the Schwinger-Keldysh timefold that computes real-time expectation values in thermal systems. This timefold contains two sides, one moving forwards in time and the other moving backwards, and they can be identified approximately with our $L,R$ systems. Preliminary study of this problem was in \cite{Endlich:2012vt,Grozdanov:2013dba,Haehl:2013hoa}, and more recently, systematic procedures for writing higher order dissipative theories of hydrodynamics in this context have been developed, starting with \cite{Crossley:2015evo,Haehl:2015uoc}. For a review see \cite{Glorioso:2018wxw}. In our case, to capture the contribution of the soft modes for very small $\omega$, we only need the simplest possible version of this theory, in 0+1 dimensions and at lowest order in derivatives. This is a non-dissipative theory, referred to as an ideal fluid.

In general, the hydrodynamic fluctuations are described by a map from the physical spacetime to the ``fluid spacetime.'' In 0+1 dimensions, this becomes simply a map from the physical time $t$ to the fluid time $\widetilde{t}$. In terms of this map, the local temperature is $\tfrac{1}{2\pi}\widetilde{t}'(t)$. At lowest order in derivatives, a Lagrangian would be some function $L(\tfrac{1}{2\pi}\widetilde{t}'(t))$. The corresponding Hamiltonian would be $E = \tfrac{1}{2\pi}\widetilde{t}'(t) L' - L$. This will indeed be the energy of the theory at temperature $\widetilde{t}'(t)$ if we choose $L = -F$ where $F$ is the free energy.\footnote{We learned this special case of the hydrodynamic actions from Juan Maldacena.} The total action for the combined $L,R$ system should then be
\be\label{hydroaction}
I_{\text{Lor}} = -\int_0^T dt \left[F(\tfrac{1}{2\pi}\widetilde{t}'_L) - F(\tfrac{1}{2\pi}\widetilde{t}'_R)\right].
\ee
The function $F$ is the free energy as a function of the temperature. Since we work in terms of the inverse temperature $\beta$, this is $F(\frac{1}{\beta})$.

Now, we would like to use this to evaluate the action for the soft modes. We consider small fluctuations about the saddle point as appropriate for a one-loop determinant:
\be
\widetilde{t} = 2\pi\frac{t + \epsilon(t)}{\beta_{\text{aux}}}.
\ee
To quadratic order in the fluctuations $\epsilon$, we have
\be\label{quadHydro}
I_{\text{Lor}} = -\tfrac{1}{2\beta_{\text{aux}}^2}F''(\tfrac{1}{\beta_{\text{aux}}})\int dt \left[(\epsilon'_L(t))^2 - (\epsilon'_R(t))^2\right],
\ee
where the primes on $F''$ refer to derivatives with respect to the temperature $1/\beta_{\text{aux}}$. This prefactor can be rewritten using thermodynamic identities as
\be
\tfrac{1}{\beta_{\text{aux}}^2}F''(\tfrac{1}{\beta_{\text{aux}}}) = \beta\frac{dE}{d\beta}\Big|_{\beta = \beta_{\text{aux}}}.
\ee
Next, the action (\ref{quadHydro}) can be translated into an action for the modes $\Deltat(t)$ and $\db_{\text{aux}}(t)$. These variables can be related to $\epsilon_L(t),\epsilon_R(t)$ by 
\be\label{setsOfVars}
\epsilon_R(t) - \epsilon_L(t) = \Deltat(t), \hspace{20pt} \frac{\epsilon_L'(t) + \epsilon_R'(t)}{2} = -\frac{\db_{\text{aux}}(t)}{\beta_{\text{aux}}}.
\ee
The first equation is clear. The second equation says that $\db_{\text{aux}}$ is the average of the fluctuation in the inverse temperature on the $L$ and $R$ systems. In terms of these variables the action is
\be\label{betadeltaaction}
I_{\text{Lor}} = -E'(\beta_{\text{aux}})\int_0^T dt \,\db_{\text{aux}}(t)\Deltat'(t)= -\int_0^T dt \,E(t)\Deltat'(t).
\ee
In the final expression we changed variables from $\delta\beta(t)$ to $E(t) = E'(\beta_{\text{aux}})\db(t)$.\footnote{More precisely, this holds for the nonzero modes. We use the notation $E$ instead of $\dE$ because it is more convenient to have the average value be the actual energy, rather than the fluctuation about some reference energy. Also, note that the transformation from $\db_{\text{aux}}$ to $E$ is ultralocal, so there will be no Jacobian in the path integral measure.} 

Before we can compute the one-loop determinant with this action, we need to choose a measure. In principle, in defining the path integral for SYK, we should use an ultralocal measure in the functional space for $G(t,t')$, such as the one associated to the pairing (\ref{pairing}) discussed below. For low frequency, the fluctuations in $E(t)$ and $\Deltat(t)$ correspond to fluctuations in e.g~$G$ given by (\ref{fluctuation}). Using this formula, one finds that the pairing for $\delta G$ induces an ultralocal metric for $E(t)$ and $\Deltat(t)$:
\be
ds^2 = \big(d(\delta G),d(\delta G)\big) = \frac{(\partial_\Deltat G,\partial_\Deltat G)}{T}\int_0^T dt (d\Deltat(t))^2 + \frac{(\partial_E G,\partial_E G)}{T}\int_0^T dt (dE(t))^2,
\ee
where we used that $(\partial_\Deltat G,\partial_E G) = 0$. We can rescale this metric by an ultralocal transformation that removes the factors of $(\partial_\Deltat G,\partial_\Deltat G)$ and $(\partial_E G,\partial_E G)$, so that the metric we use for e.g~$\Deltat(t)$ is
\be
ds^2 = \frac{1}{T}\int_0^T dt\,(d\Deltat(t))^2 = d\Deltat_0^2 + 2\sum_{n=1}^\infty \left[ (d\Deltat_n^{(R)})^2 + (d\Deltat_n^{(I)})^2\right].
\ee
On the RHS we are working in Fourier space
\begin{align}
\Deltat(t) &= \sum_{n = -\infty}^\infty e^{-i\omega_n t} \Deltat_n, \hspace{20pt} \Deltat_n = \Deltat_{-n}^*= \Deltat_n^{(R)} + i\Deltat_n^{(I)}.
\end{align}
The corresponding path integral measure is
\be
\mathcal{D}\Deltat(t) = d\Deltat_0 \prod_{n=1}^\infty 2d\Deltat_n^{(R)}d\Deltat_n^{(I)}
\ee
and similarly for $\mathcal{D}E(t)$.\footnote{This discussion of the measure might have felt overly elaborate; the only point is that we should use an ultralocal measure in the space of $E(t),\Deltat(t)$ and not in the space of $\epsilon_L(t),\epsilon_R(t)$. These are different because the sets of variables are related by a differential equation (\ref{setsOfVars}).}

Finally, let's compute the path integral. Writing the action (\ref{betadeltaaction}) in terms of the modes $\Deltat_n$ and $E_n$, we find
\be\label{hydroActionFourier}
I_{\text{Lor}} = 2T \sum_{n = 1}^\infty\omega_n\left[E_n^{(I)}\Deltat_n^{(R)}-E_n^{(R)}\Deltat_n^{(I)} \right].
\ee
Using $\omega_n = \frac{2\pi n}{T}$ and doing the Gaussian integral over the nonzero modes $(n\ge 1)$ gives
\be
\int \mathcal{D}\Deltat(t)\mathcal{D}E(t)e^{iI_{\text{Lor}}} = \int d\Deltat_0\, dE_0\prod_{n = 1}^\infty\frac{1}{n^2}.
\ee
The infinite product can be regularized by zeta function regularization or a more physical smooth cutoff, and we find
\be
-2\sum_{n=1}^\infty \log(n)\rightarrow (\text{local div.})+\log\left(\frac{1}{2\pi}\right).
\ee
The local term is UV divergent, but it would be finite in e.g.~the exact SYK theory, where the hydrodynamic description is cut off at short distances. In any case this term is proportional to $T$, and it must combine with other terms proportional to $T$ coming from the generic modes discussed above to give zero contribution. This justifies keeping only the finite piece. So finally
\be\label{finalResult}
\text{(1-loop det.)} = \int \mathcal{D}\Deltat(t)\mathcal{D}E(t)e^{iI_{\text{Lor}}} = \int \frac{d\Deltat_0 dE_0}{2\pi}= 2T\int \frac{dE_{\text{aux}}}{2\pi}.
\ee
In the final expression we used that the exact zero mode $\Deltat_0 = \Deltat$ runs from zero to $2T$, since it describes the relative time shift between two time coordinates that are defined on a circle of size $T$ with antiperiodic boundary conditions. For a theory without fermions, we wouldn't have the factor of two.

\subsection{Factors of two in the application to SYK}
\subsubsection{\texorpdfstring{$\mathbb{Z}_2$}{Z\_2} or \texorpdfstring{$\mathbb{Z}_4$}{Z\_4} orbits of saddle points}
The result (\ref{finalResult}) is almost the final answer for SYK, but not quite. This is because there are actually multiple saddle points for each value of the two zero modes $\beta_{\text{aux}}$ (or equivalently $E_{\text{aux}}$) and $\Deltat$. This can be explained by symmetries of the saddle point equations for the two point function $G_{ij}(t,t')$. The equations are
\begin{align}\label{verifying}
\partial_{t_1}G_{i_1i_3}(t_1,t_3) - \sum_{i_2}\int_0^T dt_2\Sigma_{i_1i_2}(t_1,t_2) G_{i_2i_3}(t_2,t_3) &= \delta(t_{1}-t_{3})\delta_{i_1i_3}\\\Sigma_{ij}(t,t') &= s_{ij}J^2G_{ij}^{q-1}(t,t')
\end{align}
where $s_{LL} = s_{RR} = -1$ and $s_{LR} = s_{RL} = (-1)^\frac{q}{2}$.

For any even value of $q$, these equations have a symmetry under $G_{LR}(t,t')\rightarrow -G_{LR}(t,t')$, in the sense that if we have a a solution, then after acting with this transformation we will still have a solution. This transformation can be understood as acting with $(-1)^F$ on the $R$ system but not the $L$ system. This symmetry acts nontrivially on the ``ramp'' saddle points discussed in section \ref{secbeta0} for each value of $\beta_{\text{aux}}$ (or $E_{\text{aux}}$) and $\Deltat$. However, it can be understood as simply advancing $\Deltat$ by $T$, and since we integrate $\Deltat$ from zero to $2T$, we are already integrating over all of the solutions in (\ref{finalResult}). Equivalently, we could have integrated $\Deltat$ from zero to $T$, and then multiplied by two because of this $(-1)^F$ symmetry.

\begin{figure}[t]
\begin{center}
\includegraphics[width=.45\textwidth]{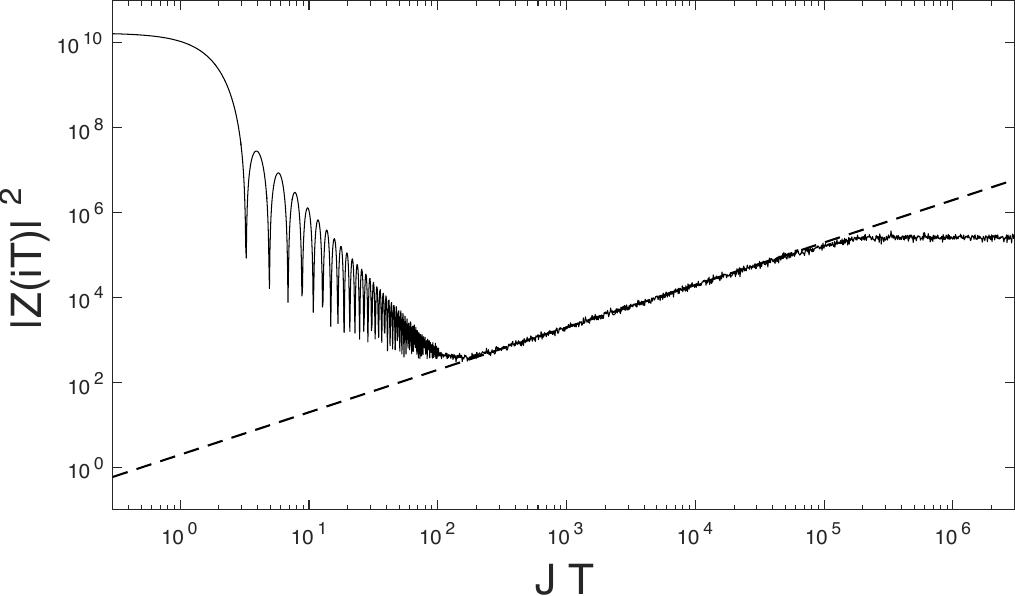}
\hspace{25pt}
\includegraphics[width=.4\textwidth]{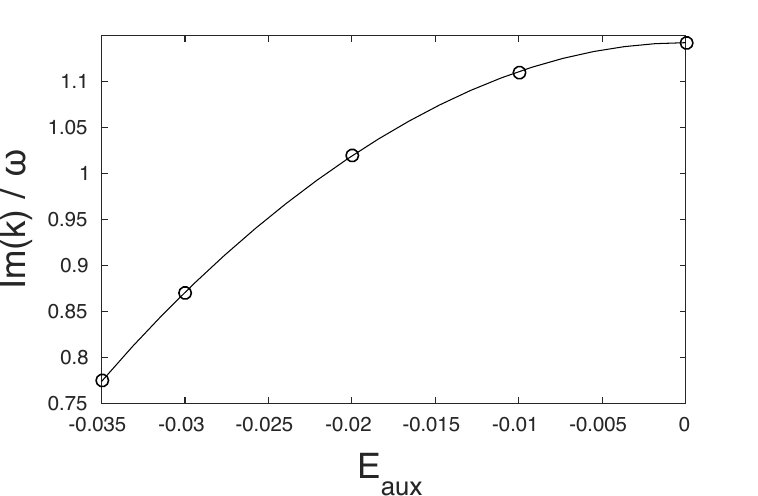}
\caption{{\small{\bf Left:} the solid curve is numerical data from \cite{Cotler:2016fpe} for $|Z(iT)|^2$  in SYK with $N = 34$ and $q = 4$, averaged over 90 realizations. The straight dashed line is $4\cdot\frac{E_{max}-E_{min}}{2\pi}\cdot T$, see (\ref{q4}). In the plot, the value $E_{max}-E_{min} \approx 3.135 \,J$ was taken directly from the numerical spectra. {\bf Right:} the solid curve represents the prediction for $\text{Im}[k(\omega)]/\omega$ (see section \ref{checking}) for small $\omega$, based on the hydrodynamic action. The circles represent a direct numerical computation of the kernel.}}\label{fig:comparetoED}
\end{center}
\end{figure}

When $q$ is a multiple of four, one can check that a ``square root'' of this transformation is also a symmetry. This transformation leaves $G_{LL}$ alone and acts as
\be
G_{LR}(t,t')\rightarrow i G_{LR}(t,-t'), \hspace{10pt} G_{RL}(t,t')\rightarrow i G_{RL}(-t,t'), \hspace{10pt} G_{RR}(t,t') \rightarrow -G_{RR}(-t,-t')
\ee
on the other components. The reason that $q$ must be a multiple of four is that in this case, the above transformation flips the relative sign between $G_{LR}$ and $\Sigma_{LR}$, which is important for verifying that (\ref{verifying}) still holds. This transformation is a type of time-reversal acting only on the $R$ system: $\psi_R(t)\rightarrow i \psi_R(-t)$.\footnote{This seems reasonable since SYK is time-reversal invariant only when $q$ is a multiple of four. We can't be more precise because we don't know how to define time reversal for subsystems (even noninteracting ones) in cases where the state of the full system is entangled.} It acts nontrivially on the saddle points of section \ref{secbeta0}, and it forms a $\mathbb{Z}_4$ group that contains the previous symmetry as a subgroup. So we get an extra factor of two, as reported in (\ref{q4}).\footnote{One is tempted to say that complex conjugation is another symmetry of these equations, and since the solutions are in general complex, this would give another $\mathbb{Z}_2$ orbit and a corresponding factor of two. In a sense, this is true, but the correct interpretation of complex conjugation of $G_{ij}(t,t')$ is that it changes the energy $E_{\text{aux}}$ to $-E_{\text{aux}}$. So, such solutions are already counted when we integrate over $\int dE_{\text{aux}}$ including both positive and negative values.} This formula is compared to exact diagonalization data in figure \ref{fig:comparetoED}. Although we won't plot it, we also checked (\ref{q6}) using $q = 6$ data from \cite{masaki}, and the agreement is similarly good.

\subsubsection{RMT expectations}\label{app:rmtExpectations}
One can also compare (\ref{q6}) and (\ref{q4}) to random matrix theory expectations. As a first step, we need to know the normalization of the ramp in the pure GOE, GUE, and GSE ensembles. This is derived rigorously in \cite{mehta2004random}, but it can be derived in a quick and dirty way following \cite{altshuler1986repulsion,Cotler:2016fpe}. For GOE, GUE, GSE, the joint distribution for the eigenvalues is
\be\label{RMTbeta1}
P(\{\lambda_i\})\propto \prod_{i<j}|\lambda_i-\lambda_j|^{\b} e^{-\frac{L\b}{4}\sum_i \lambda_i^2} \hspace{20pt}
\ee
where the coefficient $L\b/4$ corresponds to a particular choice of normalization of the eigenvalues, and $\b = 1,2,4$ for GOE, GUE, GSE, respectively.\footnote{We use boldface $\b$ for the parameter in (\ref{RMTbeta1}), to distinguish from the inverse temperature $\beta$ elsewhere.} The important difference between ensembles is that the first (Vandermonde determinant) factor is raised to the power $\b$. In terms of the distribution $\rho(\lambda) = \frac{1}{L}\sum_i \delta(\lambda - \lambda_i)$ we can write
\be
P[\rho(\lambda)] \propto \exp\left\{-\frac{L^2\b}{4}\int d\lambda\, \rho(\lambda)\lambda^2 + \frac{\b L^2}{2}\int\int d\lambda d\lambda' \,\rho(\lambda)\rho(\lambda')\log|\lambda-\lambda'|\right\}.
\ee
The saddle point of this action is the Wigner semicircle. Expanding about the saddle point $\rho(\lambda) = \rho_{sc}(\lambda) + \delta\rho(\lambda)$, we get a quadratic action from the second term in the action. It is clear that the two point function of $\delta\rho$ will be proportional to $1/\b$. By working out the propagator explicitly, see (39) of \cite{Cotler:2016fpe}, one can show that 
\be\label{RMTbeta}
\langle |Z(iT)|^2\rangle \supset L^2\int d\lambda d\lambda' e^{iT(\lambda-\lambda')}\langle\delta \rho(\lambda)\delta\rho(\lambda')\rangle =  \frac{T}{\pi\b}\int d\overline{\lambda} =  \frac{T}{\pi\b}(\lambda_{max} - \lambda_{min}).
\ee
where $\overline{\lambda}$ is the sum of the two eigenvalues, and we have done the integral over the difference. The coefficient of $T$ here agrees with the coefficient of $T$ derived more systematically in \cite{mehta2004random}, where corrections in powers of $T/L$ for the GOE and GSE cases are also derived.

Let's now apply this to SYK in the case $q \equiv 0$ (mod 4). We consider separately the cases where the statistics are GOE, GUE, and GSE, which depends on $N$ mod 8 \cite{You:2016ldz} (see table 2 of \cite{Kanazawa:2017dpd} for a nice summary). First, for the values of $N$ where we get GOE statistics, the Hamiltonian consists of two independent blocks, acting on even and odd fermion parity states. Assuming GOE statistics for each block, we get twice the GOE answer ($\b = 1$) from (\ref{RMTbeta}), which agrees with (\ref{q4}). Second, for the values of $N$ where we get GUE statistics, the two fermion parity sectors are related to each other by time reversal, and are therefore degenerate. So the total spectrum consists of a two-fold degenerate set of levels in the GUE class. Two-fold degeneracy in a set of eigenvalues leads to a factor of four in the spectral form factor, so we expect to get four times the GUE answer ($\b  =2)$ in (\ref{RMTbeta}), again consistent with (\ref{q4}). Finally, for the values of $N$ where we get GSE statistics, the two blocks are again independent, as for GOE, but now they are each two-fold degenerate. This leads to $2\cdot 4 = 8$ times the naive GSE answer, where the two is for the two blocks, and the four is for the degeneracy within each block. Using (\ref{RMTbeta}) with $\b = 4$, we get consistency with (\ref{q4}).

In the case $q \equiv 2$ (mod 4), there is never any degeneracy, and the relevant RMT ensembles always have $\b = 2$, see table 3 from \cite{Kanazawa:2017dpd}. So we expect to get twice the $\b = 2$ answer, where the factor of two comes from the two blocks. This agrees with (\ref{q6}).

\subsection{Checking the hydrodynamic action}\label{checking}
The ideal fluid action (\ref{hydroaction}) played a central role in this appendix. In addition to the general arguments for its validity, we can give some direct evidence for this in SYK. First, as pointed out in \cite{Jensen:2016pah}, the Schwarzian action contains a version of this hydrodynamic theory. At leading order in the low-frequency expansion, we discard the $\widetilde{t}''$ terms in the Schwarzian action. What remains is (\ref{hydroaction}). So the hydrodynamic action is correct in the low energy region where the Schwarzian theory applies.

For higher energies, we can check the action numerically. We can parametrize the soft modes by a fluctuation in $\delta G$. After integrating out $\delta \Sigma$, the quadratic action for $\delta G$ is (see section 4 of \cite{Maldacena:2016hyu})
\be\label{kernelAction}
I_{\text{Euc}} = \frac{N(q{-}1)J^2}{4}g\cdot (\hat{K}^{-1}-1)g, \hspace{20pt} g_{ij}(t,t') \equiv \delta G_{ij}(t,t')\widetilde{G}_{ij}(t,t')^{\frac{q-2}{2}}.
\ee
where $\widetilde{G}$ is defined in (\ref{widetildeG}). The dot product $g^{(1)}\cdot g^{(2)}$ for our Lorentzian two-contour case is defined as
\be
g^{(1)}\cdot g^{(2)} = - \sum_{i,j = L,R}\int dt dt' g_{ij}^{(1)}(t,t')g^{(2)}_{ij}(t,t').
\ee
and the corresponding pairing for perturbations $\delta G$ is
\be\label{pairing}
\left(\delta G^{(1)},\delta G^{(2)}\right) = g^{(1)}\cdot g^{(2)}= \sum_{i,j=L,R}\int dt dt's_{ij}\delta G_{ij}^{(1)}\delta G_{ij}^{(2)}G_{ij}^{q-2}.
\ee
As always, $s_{LL} = s_{RR} = -1$ and $s_{LR} = s_{RL} = (-1)^\frac{q}{2}$. The kernel can be partially diagonalized by going to frequency space for the sum of the times, as in (\ref{partialDiag}). By solving the saddle point equations numerically and discretizing this reduced kernel as a matrix, one can find the eigenvalues explicitly. For small frequency, and in the antisymmetric sector, one finds that there are a pair of eigenvalues that are close to one, with the approximate form
\be
k(\omega_n) = 1 \pm i \omega_n c
\ee
for a constant $c$ that depends on $E_{\text{aux}}$ and can be read off from the numerical results.

We would like to check that the hydrodynamic action (\ref{betadeltaaction}) equals the action (\ref{kernelAction}) with 
\be\label{fluctuation}
\delta G(t_1,t_2) = \partial_\Deltat G(t_1{-}t_2)\sum_{n} \Deltat_n e^{\frac{i}{2}\omega_n(t_1+t_2)}\, + \,\partial_E G(t_1{-}t_2)\sum_{n}E_n e^{\frac{i}{2}\omega_n(t_1+t_2)}.
\ee
Let's focus on a particular frequency pair $\pm\omega_n$ with $n\ge 1$. Then we have a four-dimensional subspace of perturbations $\delta G$, corresponding to $\Deltat^{(R)}_n,\Deltat^{(I)}_n,E^{(R)}_n,E^{(I)}_n$. The prediction for the action as a function of these parameters is (\ref{hydroActionFourier}). It is hard to compare this directly to the action (\ref{kernelAction}), since in general we expect the eigenvectors of $K$ to correspond to linear combinations of these parameters. However, we can compare the determinant of the quadratic form that defines the action. If we let $x_a$ denote any basis for this four dimensional space, then we can compute a basis-independent quantity
\be
\frac{\det\partial_{x_a}\partial_{x_b} I}{\det \big(\partial_{x_a}\delta G,\partial_{x_b}\delta G\big)}.
\ee
Requiring that this be equal for {\it (i)} the basis $\Deltat^{(R)}_n,\Deltat^{(I)}_n,E^{(R)}_n,E^{(I)}_n$ using the hydrodynamic action (\ref{hydroActionFourier}) and {\it (ii)} the basis that corresponds to the real and imaginary parts of the orthonormal eigenvectors of the kernel using the action (\ref{kernelAction}) and expanding to leading order in $\omega_n$, we find
\be
\frac{(T\omega_n)^4}{\big(\partial_E G,\partial_E G\big)^2\big(\partial_\Deltat G,\partial_\Deltat G\big)^2} = \left(\frac{N(q{-}1)J^2}{2}\right)^4 (\omega_n c)^4.
\ee
This equation can be solved to give $c$ in terms of the norms of $\partial_E G$ and $\partial_\Deltat G$, which can be computed by numerically solving the saddle point equations to obtain $G,\Sigma$ and their derivatives. So the hydrodynamic action gives a prediction for $c= \text{Im}[k(\omega_n)]/\omega_n$ in terms of these quantities. We plot this as a function of $E_{\text{aux}}$ in the right panel of figure \ref{fig:comparetoED}. At several energies we compare to the actual expression computed by diagonalizing the kernel numerically. The agreement is good.

\subsection{Exact Schwarzian version}\label{JTexact}
So far in this appendix, we only computed a one-loop determinant. At low energies or long times, one might be worried about large corrections coming from the quantum Schwarzian theory. Because the double-cone saddle point breaks $SL(2,\mathbb{R})$, the Schwarzian theory that describes the fluctuations about the low-energy saddle point is somewhat different from the standard one that describes the Euclidean partition function. Happily, it turns out that this theory is still one-loop exact. (Even better, it doesn't seem to matter much because quantum effects for large $T$ are small even by naive power counting.)

In the JT gravity variables, the problem we are interested in is to integrate over the wiggly boundary in figure \ref{figdoubleCone}, holding fixed the periodicity $\T$ of the double cone. The action for the two boundaries is the Schwarzian action, which we wrote in SYK-like variables in (\ref{fullaction}). Here e.g. $f_L = \tanh(\frac{\t_L}{2}\sqrt{1+\frac{\beta^2}{T^2}})$, where $\t_L$ is a coordinate in the the rigid double cone metric (\ref{metric}). The wiggly boundary on the left is determined by giving $\t_L$ as a function of the physical (proper) time along the boundary, $t$. The action for the left boundary is:
\be
I_{Sch} = \frac{\alpha_S}{\mathcal{J}(\frac{\beta}{T}+i)}\int_0^Tdt\left[\frac{1}{2}\left(\frac{\t_L''}{\t_L'}\right)^2 + \frac{1+\frac{\beta^2}{T^2}}{2}(\t_L')^2\right].
\ee
The action for the right boundary is the same with $i \rightarrow -i$ and $\t_L\rightarrow \t_R$. Here $\t$ is periodic, $\t \sim \t + \T$. It is convenient to define a rescaled variable $\phi = \frac{2\pi}{\T}\t$ and a rescaled time variable $u = \frac{2\pi}{T}t$. Then the action is
\be
I_{Sch} = \frac{2\pi\alpha_S}{\mathcal{J}(\beta+iT)}\int_0^{2\pi}du\left[\frac{1}{2}\left(\frac{\phi_L''}{\phi_L'}\right)^2 + \frac{1+\frac{\beta^2}{T^2}}{2}\left(\frac{\T}{2\pi}\right)^2(\phi_L')^2\right].
\ee
This is an example of the type of action discussed in section 3.1 of \cite{Stanford:2017thb}, where $b_0 \neq -\frac{c}{24}$. The action has a $U(1)$ symmetry, but not $SL(2,\mathbb{R})$ as in the usual Schwarzian case. In the bulk this is because the double cone only has a $U(1)$ isometry. This theory is one-loop exact, so we do not need to worry about quantum Schwarzian corrections to the one-loop analysis in this appendix. However, note that even without this one-loop exactness, for large $T$, loop corrections in this theory are already small by naive power counting. Concretely, if we write $\phi_L(u) = u + \tfrac{1}{\sqrt{T}}\varepsilon(u)$, then the propagator for $\varepsilon$ will be independent of $T$ for large $T$, and the interaction terms will come with inverse powers of $T$.  Here $T$ and $\beta_{\text{aux}}$  appear in the ratio $\T = T/\beta_{\text{aux}}$, and so large $\beta_{\text{aux}}$ does not suppress interaction corrections by power counting.   This is different than the usual Schwarzian case, where  neither large  $T$ nor large $\beta_{\text{aux}}$  suppress interaction effects by power counting, although the dynamics is again one loop exact.   This difference reflects the fact that on the ramp large $T$ probes small energy differences, while on the slope large $T$ probes small energies.

{\small
\bibliography{references}
}

\bibliographystyle{utphys}

\end{document}